\documentclass[conference]{IEEEtran}
\IEEEoverridecommandlockouts
\usepackage{cite}
\usepackage{amsmath,amssymb,amsfonts}
\usepackage{algorithmic}
\usepackage{graphicx}
\usepackage{textcomp}
\usepackage{xcolor}
\usepackage{amsthm}
\usepackage{hyperref}
\usepackage{adjustbox}
\graphicspath{{new_figures/}}
\usepackage[linesnumbered,ruled,vlined]{algorithm2e}
\usepackage{comment}
\usepackage{booktabs}

\newcommand{\tool}{CSD-JWT}
\usepackage{adjustbox}
\usepackage{makecell}
\usepackage{amssymb}
\usepackage{amsmath}
\usepackage{subcaption}
\newtheorem{definition}{Definition}
\definecolor{pastelgray}{rgb}{0.86, 0.86, 0.86}
\newcommand{\review}[1]{\textcolor{black}{#1}}
\newcommand{\asia}[1]{\textcolor{black}{#1}}

\newcommand{\ale}[1]{\textcolor{black}{#1}}
\newcommand{\car}[1]{\textcolor{black}{#1}}

\usepackage{fancyhdr}
\IEEEoverridecommandlockouts
\makeatletter
\def\ps@IEEEtitlepagestyle{
  \def\@oddhead{\begin{minipage}{\textwidth}
    \centering
    This work has been accepted to the 2025 Annual Computer Security Applications Conference (ACSAC)\\
    \rule{\textwidth}{0.2pt}
  \end{minipage}}
}

\makeatother
\usepackage[compatibility=false]{caption}
\begin{document}

\title{\title{Compact and Selective Disclosure for Verifiable Credentials}
}

% \author{
% \IEEEauthorblockN{Anonymous Author(s)}
% }

\author{
    \IEEEauthorblockN{Alessandro Buldini\IEEEauthorrefmark{1}, Carlo Mazzocca\IEEEauthorrefmark{2}, Rebecca Montanari\IEEEauthorrefmark{1}, Selcuk Uluagac\IEEEauthorrefmark{3}}
    \IEEEauthorblockA{\IEEEauthorrefmark{1}University of Bologna, Bologna, Italy}
    \IEEEauthorblockA{\IEEEauthorrefmark{2}University of Salerno, Fisciano, Italy}
    \IEEEauthorblockA{\IEEEauthorrefmark{3}Florida International University, Miami, Florida, USA}
    \IEEEauthorblockA{Email: \{alessandro.buldini,rebecca.montanari\}@unibo.it, cmazzocca@unisa.it, suluagac@fiu.edu}
}
\maketitle

\begin{abstract}
Self-Sovereign Identity (SSI) is a novel \asia{identity model} that \asia{empowers} individuals \asia{with} full control over their data, \asia{enabling} them to choose \emph{what} information to disclose, \emph{with whom}, and \emph{when}.
This paradigm is rapidly gaining traction worldwide, supported by numerous initiatives such as the European Digital Identity (EUDI) Regulation or Singapore's National Digital Identity (NDI). For instance, by 2026, the EUDI Regulation will enable all European citizens to seamlessly access services across Europe using Verifiable Credentials (VCs).
A key feature of SSI is the ability to selectively disclose only \asia{specific claims within a credential, enhancing the privacy protection of the identity owner.} This paper proposes a novel mechanism designed to achieve \textit{Compact} and \textit{Selective Disclosure} for VCs (\tool{}). Our method leverages a cryptographic accumulator to \asia{encode} claims within a credential into a unique, \asia{compact representation}. We implemented \tool{} as an open-source solution and extensively evaluated its performance under various conditions. \tool{} \asia{provides significant memory savings, lowering usage by up to 46\% compared to the state-of-the-art. It also minimizes network overhead \asia{by producing} remarkably smaller Verifiable Presentations (VPs), with size reduction from 27\% to 93\%.}
Such features make \tool{} especially well-suited for resource-constrained devices, \asia{including hardware wallets designed for managing credentials.}
\end{abstract}


\section{Introduction}
\asia{The growing emphasis on preserving user privacy has driven a significant transformation in the digital identity ecosystem, shifting from centralized to decentralized models \cite{mazzocca2024survey}. Emerging privacy-preserving regulations further reinforce this evolution \cite{gdpr2016,californiaccpa,eu2024regulation1183} that place individuals at the center of the identity management process. This paradigm shift has given rise to \emph{Self-Sovereign Identity} (SSI) \cite{MUHLE201880}, a novel identity model that is gaining worldwide traction. In February 2024, the European Commission approved the "electronic IDentification, Authentication, and trust Services" (eIDAS 2.0) \cite{eidas2.0}, mandating the creation of the European Digital Identity Wallet (EUDI Wallet) \cite{wallet}. Under this framework, by 2026, European member states must provide citizens with a digital identity wallet for managing their credentials. Similarly, Singapore's National Digital Identity (NDI) system \cite{singaporeNDI}, implemented through Singpass, enables citizens to seamlessly access public and private services while maintaining full ownership of personal data.}

The \asia{core} principle \asia{of SSI} consists of empowering individuals with full control over their data \asia{by allowing them to
decide \emph{what} information to disclose, \emph{with whom}, and \emph{when}}, \asia{eliminating reliance} on centralized authorities. This is achieved through \asia{Verifiable Credentials (VCs) \cite{krul2024sok}}, digital credentials \asia{that} attest \asia{specific} properties or attributes of the \asia{holder}, hereinafter called \emph{claims}. \asia{Unlike traditional Public Key Infrastructure (PKI) certificates \cite{pki_survey}, which 
provide secure public key distribution and rely on centralized authorities, VCs 
are designed for broader use cases and operate within a decentralized framework. These credentials} are cryptographically signed by trusted issuers (e.g., governments), \asia{enabling decentralized verification} of the \asia{stored} information.
\asia{A key feature of SSI is the ability for identity owners to \textit{selectively disclose} only a specific subset of claims contained in VCs. For instance, individuals can prove a legal drinking age without revealing unnecessary information like their address.}

\asia{Various VC formats have been proposed \cite{format_vc}, with JSON Web Token (JWT) \cite{jones2015rfc} emerging as one of the most widely adopted, as it offers a concise approach to exchanging verifiable claims.}
To \asia{enable} privacy-preserving selective disclosure, the Internet Engineering Task Force (IETF) proposed \asia{the} Selective Disclosure for JWTs (SD-JWT) \cite{SD-JWT} \asia{specification}. SD-JWT replaces plain text claims with digests of their salted values. \asia{The issuer signs the resulting JWT and shares it with the holder, along with the original claims and salts. To disclose data, the holder provides the verifier with the signed JWT and salt-claim pairs corresponding to disclosed claims.} 

\smallskip
\noindent \textbf{Problem Definition.} \asia{A major downside of this method is that credential size} grows linearly with the number of claims, \asia{increasing} end-user storage \asia{and network} requirements. Moreover, although digests \asia{hide undisclosed} information, SD-JWT reveals the exact number of included \asia{claims, raising privacy concerns. This knowledge can potentially expose the identity owner} to \textit{inference attacks} \cite{mehnaz2022your, 8068648} \asia{as the number of claims may be exploited to infer details about hidden information.} 

These drawbacks become especially critical when VCs \asia{are applied to} entities beyond individuals, such as Internet of Things (IoT) devices \cite{mazzoccaevoke, fotiu,mahalle2020rethinking}.  
\asia{Given the heterogeneous capabilities of these devices, considerations around storage, computation, and networking are even more crucial. In constrained environments, resources are limited and shared across different functions. For example, memory must store both VCs and firmware, making even KB-level reductions impactful. Moreover, minimizing resource usage is essential for enhancing energy efficiency, a critical factor in resource-constrained settings. }
\asia{These concerns also apply to} 
hardware wallets, \asia{known for their} enhanced security \asia{in managing} VCs \cite{ccshardware, 9122595}, \asia{which often} have limited storage capacities ranging from a few KB to a couple of MB \cite{ledgermemory}. 
\car{This raises the research question: \emph{How can selective disclosure mechanisms balance computation, storage, network overhead, and privacy in resource-constrained environments?}
}

\smallskip
\noindent \textbf{\tool{}.} \asia{This paper proposes a protocol that achieves Compact and Selective Disclosure for VCs (\tool{}), diminishing storage and network overhead of SD-JWT}. \asia{Our method gives identity owners} full control over their information \asia{while} minimizing the amount of data to store and share. \tool{} \asia{replaces the full list of salted hashes with a fixed-length value}, \car{obtained by aggregating} all claims into a cryptographic \ale{trapdoor-based} accumulator \cite{benaloh_demare}. \car{This significantly reduces the credential size: rather than storing a separate hash and salt for each claim, \tool{} requires only a single accumulator value for the entire credential and one \emph{witness} per claim, i.e., a proof of inclusion.}

\car{It is worth noting that the design of} \ale{ \tool{} is agnostic of the specific accumulator implementation. \car{This flexibility allows the underlying accumulator to be replaced as more secure, efficient, or space-optimized constructions emerge.} The only essential requirement for the accumulator employed in \tool{} is that it must be trapdoor-based.}
Furthermore, \ale{with the substitution of the signed list of hashes with a single accumulator value, \tool{} allows to hide} the exact number of claims in the VC, offering stronger resilience against inference attacks. \car{In contrast, SD-JWT requires the use of decoy digests to achieve similar privacy guarantees, which further increases the overall size of the credential \cite{ietf-oauth-selective-disclosure-jwt-22}}.

\car{In \tool{},} the issuer must provide a witness for each claim to be disclosed. The holder then combines these witnesses with corresponding claims to generate Verifiable Presentations (VPs). The storage requirements for disclosing claims, i.e., \car{accumulator value and witnesses, are lower than \asia{those} for signed list} and salts due to the constant size of \car{the accumulator value.}
This property also \asia{reduces} the amount of data \asia{shared} with the verifier, making \tool{} \asia{an efficient} solution for constrained environments.

\smallskip
\noindent \textbf{Evaluation.} We implemented a prototype of \tool{} \ale{based on the cryptographic accumulator proposed in \cite{vitto_biryukov},} and conducted a series of experiments, \asia{including tests on an RFC7228 class-2 constrained device \cite{rfc7228},} to compare \asia{our method with} the state-of-the-art mechanism, i.e., SD-JWT. 
\review{\asia{In particular,} we measured storage \asia{and computational} requirement to evaluate \tool{} suitability for constrained devices.} \asia{It is worth noting that our evaluation considers a holder who maintains a single credential. However, as underscored by the EUDI Regulation \cite{eu2024regulation1183}, users are likely to use multiple VCs by 2026 in real-world scenarios \cite{gartner2024digitalidentity}, making the overhead reduction offered by \tool{} even more impactful.} Experimental findings proved that \asia{our approach} can achieve up to 46\% \asia{memory} savings for \asia{each stored credential}. \asia{Experiments on} the network overhead \asia{shows that} VP in \tool{} is \asia{93\%} smaller when the holder aims to disclose only 1 claim out of 100 (maximum privacy), while it still provides a 27\% reduction when no privacy preservation is required and all claims are revealed.

\smallskip
\noindent \textbf{Contributions.} 
In the following, we summarize the key contributions of this work:

\begin{itemize}
    \item We propose \tool{}, a mechanism for compact and selective disclosure of claims within VCs, providing the holder with full control over their data;
    \item In \tool{}, all claims are mapped into a \asia{fixed-length value, concealing} the number of claims within the VC. This contributes to enhancing the privacy level as it minimizes the amount of information leaked;
    \item \tool{} remarkably reduces the amount of information to maintain and share, thereby lowering both storage and network overhead. This makes it suitable for constrained environments, such as hardware wallets \review{and} IoT devices;
    \item We implemented \tool{}, and conducted experiments to evaluate and compare the proposed mechanism with SD-JWT. We also made our code fully available to the research community. 
\end{itemize}

\smallskip
\noindent \textbf{Organization.} The remainder of this paper is structured as follows. Section \ref{sec:background} provides the background on VCs and cryptographic accumulators. Section \ref{sec:model} presents the reference system and threat model. Section \ref{sec:protocol} details how \tool{} achieves compact and selective disclosure for VCs. Section \ref{sec:security} analyzes the security of the proposed mechanism in addressing the identified threats. Section \ref{sec:evaluation} comprehensively evaluates \tool{}. Section \ref{sec:related} reviews literature in the field. Finally, Section \ref{sec:conclusion} draws our conclusions.

\section{Background}\label{sec:background}

\subsection{Verifiable Credential}
A digital identity can be envisioned as a set of attributes or properties that provides a snapshot of any entity, including individuals, companies, and objects. \asia{Any third party must be able to directly verify these attributes through attestations,} embedded within digital credentials issued by trusted authorities. 
A data structure that satisfies this property is referred to as a VC. Several VC data models have been proposed in recent years, among which the  World Wide Web Consortium (W3C) VCDM \cite{w3c_dm} and SD-JWT VC \cite{sd_jwt_vc} are considered the most prominent. We propose \tool{} as a selective disclosure mechanism that can be seamlessly integrated into any model, offering a practical alternative to SD-JWT.

To be effective and address the challenges of centralized identity models, a verifier \asia{(e.g., a service provider)} should be able to directly verify the authenticity of the claims contained, without direct interaction with the issuer. This necessitates that the issuer identity can be directly retrieved from the credential itself. Consequently, every party must have a globally unique self-generated digital identity \cite{sokdatasover}. The commonly adopted approaches are public keys and Decentralized Identifiers (DIDs) \cite{did}. A DID is a Uniform Resource Identifier (URI) standardized by the W3C. Each DID is associated with a DID Document, encompassing publicly available information (e.g.,  public key) about the entity it identifies. The DID Document is typically shared through verifiable data registries such as a blockchain \cite{10.1145/3465481.3469204}. Digital identifiers are crucial for \asia{identifying issuers and holders}, preventing the reuse of previously collected VCs, which may allow an adversary to impersonate the identity owner. In \tool{}, digital identification for both holders and issuers is achieved through DIDs. However, it could be easily extended to support public key identification.

\subsection{Cryptographic Accumulators} 

Cryptographic accumulators are cryptographic primitives that securely map a set of values in a fixed-size output called the accumulator value \cite{benaloh_demare}. Each accumulated element can be associated with a proof of inclusion, known as a witness. This enables verifying the membership without disclosing any other elements of the set. Although several implementations have been proposed, all types of accumulators must satisfy two properties: 

\begin{enumerate}
    \item \textit{One-wayness}: Given an hash function \( h_l: X_l \times Y_l \rightarrow Z_l \) this is said to be one-way if:
    \begin{itemize}
        \item For each integer \( l \), \( h_l(x,y) \) is computable in polynomial time for all \( (x, y) \in X_l \times Y_l \).
        \item Additionally, no polynomial-time algorithm exists such that, for a sufficiently large \( l \), given a pair \( (x, y) \in X_l \times Y_l \) and \( y' \in Y_l \), \review{it is possible to} find a suitable \( x' \in X_l \) that satisfies \( h_l(x, y) = h_l(x', y') \).
    \end{itemize}
    
    \item \textit{Quasi-commutativeness}: Given \( h_l: X_l \times Y_l \rightarrow X_l \), the following equality holds \( \forall x \in X_l \) and \( \forall y_1, y_2 \in Y_l \):
    \[
    h_l(h_l(x, y_1), y_2) = h_l(h_l(x, y_2), y_1).
    \]
\end{enumerate}

\noindent We apply $h_l$ to all the elements $e_i$ to be included in the accumulator, obtaining the accumulator value $a$. For each $e_i$ there exists a witness $w_i$ such that $a = h_l(e_i, w_i)$, proving its inclusion within $a$.
The one-way property ensures that given the accumulator value $a$, it is computationally infeasible to find a $w_i$ that proves membership for an $e_i$ that does not belong to the set. Moreover, using a function that meets the quasi-commutative requirement guarantees that the accumulator remains independent of the order in which values are accumulated.

Accumulators can be classified based on their support for verifying element inclusion. Specifically, those that enable verification using membership witnesses are called positive, as opposed to those supporting non-membership witnesses, which are known as negative. The ones that support both types of verification methods are called universal. 
Various accumulators have been proposed over the years \cite{camenish_lysyanskaya, nguyen, au, papamanthou, boneh_merkle_trees}. Among these, accumulators based on Bilinear Groups \cite{nguyen, au, vitto_biryukov} are widely acknowledged for their efficiency \cite{wangetal, kumaretal}, especially in terms of witness size and verification efficiency for inclusion. Given these properties, \ale{in Section \ref{sec:evaluation}, we evaluate \tool{} by instantiating it with a positive ECC-based accumulator.}

\subsection{Selective Disclosure}
In the SSI model, an identity owner or holder $h$ is any entity provided with a VC by a trusted issuer $i$. A credential comprises a set of claims $\mathcal{C}$, which are assertions made by the issuer about $h$. One of the key focuses of SSI is to empower $h$ with full control over their personal data. \asia{This is achieved through selective disclosure, which is defined as follows:}

\begin{definition}[\textbf{Selective Disclosure}]\label{def:seldisc} 
\textit{Selective disclosure is a mechanism that allows an identity owner $h$ to disclose only a subset of claims $C \subseteq \mathcal{C}$ to a verifier $v$ so that $C$ is still verifiable.}
\end{definition}

A mechanism that implements selective disclosure must grant the holder fine-grained control, ensuring that they can independently determine which claims to share in response to a verifier’s request. Simultaneously, it must adhere to the principle of data minimization by limiting the disclosure to only the information strictly necessary to satisfy the verifier’s requirements, thereby reducing the risk of excessive or unintended data exposure. Selective disclosure must also preserve the verifiability of the shared information: the verifier must be able to independently assess the authenticity and integrity of the disclosed claims without needing access to the full credential.

\section{System and Threat Model}\label{sec:model}
\begin{figure}[!t]
\centering
\includegraphics[width=0.9\columnwidth]{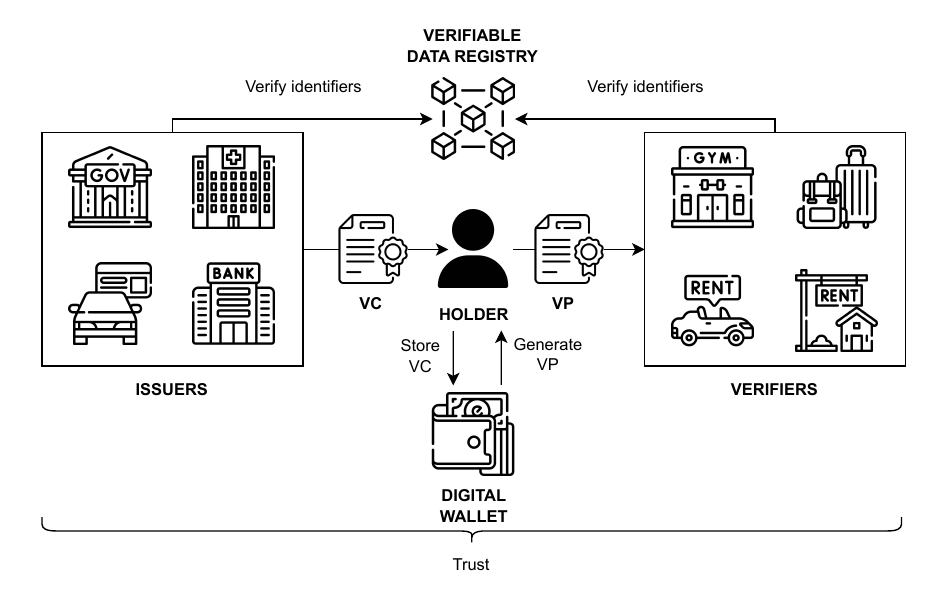}
\caption{SSI reference system.}
\label{fig:system}
\end{figure}

\subsection{Reference System}
This section describes the main actors involved in the SSI model, where all participants are digitally identified with a DID following the W3C specification. Figure \ref{fig:system} sketches the main entities and their interactions within the SSI paradigm. 

\smallskip
\noindent \textbf{Issuer.} An issuer is a trusted entity, such as a government, university, or municipality, that issues VCs to identity owners. For example, an issuer may provide an individual with a VC serving as a driving license. The issuer signs credentials with its private key, making it directly verifiable to third parties. It is worth noting that a crucial assumption of the SSI model is that holders and verifiers trust the issuer.

\smallskip
\noindent \textbf{Holder.} In the SSI paradigm, the holder, also known as the identity owner, is any entity that requires digital identification. Each holder is identified with a DID, whose corresponding DID Document is shared through a verifiable data registry. The holder collects VCs from an issuer and stores them in a credential repository such as a file system, storage vault, or hardware wallet, which securely stores and protects access to their credentials. These credentials are presented to access different services and facilities. To preserve their privacy, holders are willing to share only \asia{the} minimal amount of personal information to access a service. 

\smallskip
\noindent \textbf{Verifier.} A verifier is an entity, such as a service provider, that requests credentials from identity owners to provide a specific service. Regarding the driving license example, a car rental company may act as a verifier by requesting a digital driver's license from a customer before allowing them to rent a vehicle. The verifier checks the authenticity and validity of the VC to ensure the customer meets the driving requirements.

\smallskip
\noindent \textbf{Verifiable Data Registry.} When using DIDs as digital identifiers within VCs, it is essential to have \textit{trust anchors} that certify specific attributes associated with the identity, such as the DID Document, containing the identity owner's public key. In DID-based scenarios, this certification is achieved through verifiable data registries. These data sources, such as blockchains, are employed to share the DID Document, ensuring its immutability and auditability.

\subsection{Threat Model}
\car{In SSI, issuers are typically assumed to be trusted by all parties involved \cite{krul2024sok}, as they certify claims (of which they are inherently aware) that verifiers rely on to grant access to resources and services. Accordingly, in \tool{}, the generation of the accumulator value and corresponding witnesses is securely managed by a trusted issuer. Our threat model aligns with that of SD-JWT, which does not guarantee unlinkability in scenarios where trusted issuers collude with other entities \cite{baum2024cryptographers}. This limitation stems from the fact that hashes, salts, accumulator values, and witnesses remain constant across credential presentations, making repeated disclosures linkable under adversarial conditions. A more in-depth analysis of unlinkability and its implications can be found in Appendix~\ref{app:unlinkability}.} 
In this subsection, we provide an overview of the adversary perspective and outline the potential threats to our system, based on the threat model presented in \cite{krul2024sok}. 

\smallskip
\noindent \textbf{Adversary's Goal.} The adversary has two primary objectives: to disclose personal information within a VC and to reuse previously collected credentials to impersonate the holder and gain access to services on their behalf.

\smallskip
\noindent \textbf{Adversary's Capability.} We adopt the Dolev-Yao adversary model \cite{dolevyao}, which assumes that the adversary can eavesdrop on, intercept, and inject an unlimited number of messages.

\smallskip
\noindent \textbf{Adversary's Knowledge.} We consider a setting where the adversary is aware of the algorithm used to issue VCs and knows the selective disclosure mechanism employed by holders to present their claims. 

\smallskip
\noindent \textbf{Threats.} Our model focuses on the following threats affecting the holder while presenting their credentials. 

\begin{itemize}
    \item \textit{Replay Attacks}: An adversary may present a valid credential that belongs to another identity owner. This VC could be obtained through collusion with another holder or stolen from an unsuspecting identity owner. A service provider may wrongly grant authorization, believing claims that do not represent the adversary. 
    \item \textit{Data Overcollection}: A service provider may collect more claims than necessary to provide a service, either with malicious intent or unintentionally. Moreover, the service provider may be able to infer additional details on the identity owner, given side information such as the number of claims in the VC.
    \item \textit{Compromised Communication Channel}: An adversary may intercept a credential during interactions between two legitimate parties, potentially learning its contents or modifying it.
\end{itemize}

\section{Compact and Selective Disclosure}\label{sec:protocol}

This section introduces \tool{}, a mechanism for compact and selective disclosure of claims for VCs. \tool{} is \asia{implemented as} a JWT that leverages the features of cryptographic accumulators. Credential holders are provided with an accumulator value and a Witness-Value Container (WVC), enabling them to selectively disclose their claims while minimizing both data exposure and storage overhead. Table \ref{tab:notation} reports the adopted notation. In the remainder of this section, we first explain our use of cryptographic accumulators and then describe the main phases of our method.

\subsection{Preliminaries}

\smallskip
\noindent \textbf{Accumulator.} We aggregate all claims $\mathcal{C}$, within a verifiable credential $vc$ belonging to an identity owner $h$, into a single value $a$. The elements included in $a$ are derived by applying a hash function $h()$ to each $c_j \in \mathcal{C}$. Specifically, we use the cryptographically secure \texttt{SHA-256} hash function, which is resistant to collisions \cite{sha256}. 

\begin{table}[t]
%\scriptsize % Makes the font smaller
\centering
\caption{\tool{} notations.}
\label{tab:notation}

\begin{tabular}{ c l}
\toprule
\textbf{Symbol} & \textbf{Description} \\ \midrule
$h$ & Holder / Identity owner \\
$i$ & Issuer / Trusted entity  \\
$v$ & Verifier \\
$vc$ & Verifiable credential \\
$vp$ & Verifiable presentation \\
$(pk_h, sk_h)$ & Holder key pair \\
$(pk_i, sk_i)$ & Issuer key pair to manage the accumulator \\
$a$ & Accumulator value \\
$\mathcal{C}$ & Set of the holder claims\\
$C$ & Subset of the holder claims\\
$c_j$ & Single claim of the holder \\
$\mathcal{W}$ & Set of witnesses corresponding to $\mathcal{C}$\\
$W$ & Subset of witnesses corresponding to $C$\\
$w_j$ & Single witness corresponding to $c_j$\\
\bottomrule
\end{tabular}

\end{table}

\smallskip
\noindent \textbf{Proof of Inclusion.} For each $c_j$ included in $a$, there exists a witness $w_j$. Witnesses are pieces of cryptographic evidence that holders provide to verifiers for proving that $c_j$ is included in $a$, ensuring that it has been certified by a trusted issuer.  

\smallskip
\noindent \textbf{Functions.} In the following, we report the functions used to realize \tool{}. \car{All functions related to setup, as well as accumulator and witness generation, are performed by the issuer.}

\begin{itemize} 
    \item $par \leftarrow \mathsf{Setup}(1^\lambda)$: 
    Given the security parameter $1^\lambda$, which determines the level of security, this function initializes and outputs the accumulator parameters $par$ \ale{that comprises a key pair ($pk_i, sk_i)$ and all the necessary parameters required by the underlying cryptographic accumulator instance}.

    \item $a \leftarrow \mathsf{AccumulatorSetup}(par)$: This function receives accumulator parameters $par$ and setups the accumulator, generating the initial accumulator value $a$.
    \item $a \leftarrow \mathsf{AccumulateClaims}(a, \mathcal{C}, sk_i)$: This function takes the set of claims $\mathcal{C}$, the private key $sk_i$, \car{and aggregates the claims $c_j \in \mathcal{C}$ into the accumulator value $a$.}
    \item $\mathcal{W} \leftarrow \mathsf{ComputeWitnesses}(a, \mathcal{C}, sk_i)$: 
    Given the accumulator value $a$, the set of claims \ale{$\mathcal{C}$}, the private key $sk_i$, this function generates the set $\mathcal{W}$, where each $w_j \in \mathcal{W}$ corresponds to a $c_j \in \mathcal{C}$.
    \item $0,1 \leftarrow \mathsf{VerifyWitness}(a, c_j, w_j, pk_i)$: 
    The \car{verifier $v$ checks} whether the claim $c_j$ was accumulated in $a$. This verification requires both the witness $w_j$ corresponding to $c_j$ and the accumulator public key $pk_i$.
    \item $vp \leftarrow \mathsf{GenerateVerifiablePresentation}(a, C, W, n, sk_h)$: The identity owner $h$ uses their private key $sk_h$ to sign the accumulator value $a$, a subset of claims $C$, along with the corresponding witnesses $W$, and a nonce $n$ to prevent replay attacks, producing a verifiable presentation $vp$.
    \item $0,1 \leftarrow \mathsf{VerifyPresentation}(vp)$: 
    The verifier $v$ retrieves the public keys of $pk_h$ and $pk_i$ through their respective DID using a $\mathsf{Resolve()}$ function, which allows obtaining corresponding DID Documents. The verifier checks the presentation ownership through $\mathsf{Verify(vp,pk_h)}$. \ale{Then, it extracts the accumulator value $a$ and all claim-witness pairs $(c_j, w_j)$ from the $vp$ before calling the $\mathsf{VerifyWitness}()$ function to confirm whether the provided claims are included in $a$.} 

\end{itemize}

\subsection{Protocol}

This section describes how \tool{} enables compact and selective disclosure of claims within VCs. An identity owner or holder $h$ can selectively disclose a subset of claims $C$ among those included in a $vc$ issued by a trusted entity $i$, also known as the issuer. Beyond credentials, $i$ provides holders with the plain text of their claims $c_j \in \mathcal{C}$, along with the corresponding witnesses $w_j \in \mathcal{W}$. This information is used by $h$ to selectively disclose claims, empowering the holder with maximum control over their data, meeting the SSI requirements. 

{\scriptsize
\begin{algorithm}[t!]
\caption{Verifiable Credential Issuance}\label{alg:issuance}
\KwIn{$par, \mathcal{C}, sk_t$}
\KwOut{$vc$}
$a \gets \mathsf{AccumulatorSetup}(par)$\;
$pk_i \gets par.pk$\;
$sk_i \gets par.sk$\;
$a \gets \mathsf{AccumulateClaims}(a, \mathcal{C}, sk_i)$\;
$\mathcal{W} \gets \mathsf{ComputeWitnesses}(a, \mathcal{C}, sk_i)$\;
$wvc \gets \emptyset$\;
\ForEach{$(c_j, w_j) \in (\mathcal{C}, \mathcal{W})$}{
  $wvc \gets wvc \cup \{(c_j, w_j)\}$\;
}
$vc \gets \{a, wvc\}$\;
\end{algorithm}
}

\smallskip
\noindent \textbf{Setup.} 
The issuer securely generates the security parameter $1^\lambda$ and provides it to the $\mathsf{Setup}()$ function, producing the accumulator parameters $par$. This $par$ comprises, among other data required for the setup, a key pair $(pk_i, sk_i)$. The private key $sk_i$ is necessary to accumulate claims and produce the accumulator value, as well as to generate the witnesses, while $pk_i$ is used to verify their membership. This public key must be shared through a verifiable data registry, such as a blockchain, to ensure its authenticity and integrity. The issuer $i$ uses the $\mathsf{AccumulatorSetup}()$ function to initialize the accumulator, obtaining the initial accumulator value $a$. This setup phase occurs only once, as the generated keys can support the management of multiple credentials.

\smallskip
\noindent \textbf{Issuance of Verifiable Credential.}
The protocol begins with an identity owner $h$ requesting a $vc$ from a trusted entity $i$ (i.e., the issuer), which certifies a set of claims $\mathcal{C} = \{c_1, ..., c_n\}$. Each claim is a key-value pair, composed of the name identifying a property and its value.  
\tool{} contains an accumulator value $a$, computed by accumulating all $c_j \in \mathcal{C}$, claims also include the holder and issuer identifier, respectively $DID_h$ and $DID_i$. The use of a unique identifier ensures that two distinct $vc$ containing the same $\mathcal{C}$ do not result in the same accumulator value $a$ and set of witnesses $\mathcal{W}$. The verifier uses $pk_i$, along with witnesses $W$, to verify whether claims $C$ are included in $a$. Algorithm \ref{alg:issuance} shows how VCs are issued in \tool{}. 

The issuer executes the $\mathsf{AccumulateClaims}()$ function, providing the initial accumulator value $a$, $\mathcal{C}$, and $sk_i$, updating $a$. It is worth noting that, instead of accumulating $c_j$, we add $h(c_j)$ to the accumulator to map it as a scalar value. Thus, it is highly unlikely that two distinct claims $c_j$ and $c_k$ produce the same digest, also ensuring that each witness corresponds to a unique claim. While not explicitly mentioned, in practice, the issued $vc$ also contains all the necessary information that usually accompanies $vc$, such as the underlying cryptographic mechanisms, the credential type, and its expiry date.

Furthermore, $i$ utilizes the $\mathsf{AccumulateWitnesses}()$ function operating in batch mode, giving as input $a$, $\mathcal{C}$, and $sk_i$, and receiving as output $\mathcal{W}$, the set of witnesses corresponding to the claims included in the issued credential. In addition to $vc$, $h$ is also provided with a WVC, containing a list of claim-witness pairs necessary for selectively disclosing claims and generating VPs. 

\begin{equation}
    WVC = [(w_1, c_1), (w_2, c_2), ..., (w_n, c_n)]. 
\end{equation}

\smallskip
\noindent \textbf{Generation of Verifiable Presentations.} 
By holding $a$ and the $WVC$, $h$ may reveal all claims or selectively disclose a subset $C \subseteq \mathcal{C}$ to any verifier $v$, such as service providers. To do this, $h$ executes the $\mathsf{GenerateVerifiablePresentation}()$ function with the accumulator value $a$, the claims to disclose $C$, the corresponding witnesses $W \subseteq \mathcal{W}$, and a nonce $n$ \car{provided by $v$}, generating a verifiable presentation $vp$ signed with their private key $sk_h$. This presentation is shared as a JWT. Specifically, for each $c_j$ to be disclosed, $vp$ contains a claim-witness pair ($c_j, w_j)$. The nonce $n$ is used to prevent replay attacks, where an adversary might intercept $vp$ and reuse it to spoof $h$.

{\scriptsize
\begin{algorithm}[t!]
\caption{Verifiable Presentation Verification}\label{alg:vpv}
\KwIn{$vp$}
\KwOut{$0$ or $1$}
$pk_i \gets Resolve(vp.DID_i)$\;
$pk_h \gets Resolve(vp.DID_h)$\;

$status_{vp} \gets \mathsf{Verify}(vp, pk_h)$\;

\If{$status_{vp} = 1$}{
    \ForEach{$(c_j, w_j) \in vp$}{
        $res \gets \mathsf{VerifyWitness}(a, c_j, w_j, pk_i)$\;
        \If{$res = 0$}{
            \Return $0$\;
        }
    }
}
\Return $1$\;
\end{algorithm}
}

\smallskip
\noindent \textbf{Verification of Verifiable Presentations.} 
The verifier $v$ collects the $vp$ as JWT, including the accumulator value $a$, and the witness-value pairs $(w_j, c_j) \in WVC$ corresponding to the disclosed claims. Algorithm \ref{alg:vpv} shows how presentations are verified in \tool{}. To verify the claims presented by $h$, $v$ must be provided with the issuer's and holder's identifiers previously included in $a$, which resolve to their respective DID Documents, containing their public keys, respectively, $pk_i$ and $pk_h$. The issuer's DID Document is typically retrieved from a verifiable data registry, whereas the holder's DID Document is usually provided directly by the holder.

Thus, $v$ executes the $\mathsf{VerifyPresentation}()$ function to verify the authenticity of claims included in the presentation, and to assess ownership over the credential. After verifying the holder's signature through $pk_h$, 
the verifier proceeds by asserting that the disclosed claims $C$ are included in the presented accumulator value $a$. Then, the verifier executes the $\mathsf{VerifyWitness}()$ function with $pk_i$ for each disclosed claim $c_j \in C$ and their corresponding witnesses $w_j \in \mathcal{W}$. Both the holder's and the issuer's identifiers are claims that must be validated in this process. As done by $i$ when computing $a$, before being combined with $w_j$ to prove inclusion in $a$, each $c_j$ is hashed with $h()$ using the same algorithm employed by $i$. If the combination of all the $h(c_j)$ and $w_j$ matches $a$, the set of disclosed $C$ is authentic, and $v$ can trust the information shared.

\subsection{Discussion}
\tool{} enables the generation of VCs, requiring minimal storage, regardless of the number of hidden claims within the credential. This key feature makes \tool{} ideal for use in constrained scenarios where identity holders may have limited storage and computing capabilities, \review{e.g., when using a hardware wallet for managing VCs}. In the remainder of this subsection, we provide further considerations on this key property.

\smallskip
\noindent \textbf{Storage Requirement.}
\review{For each claim in the VC, SD-JWT requires storing a 256-bit digest and a salt whose minimum length is 128 bits. In contrast, \ale{as shown in Section \ref{sec:evaluation}}, by \ale{implementing} the ECC-based accumulator proposed in \cite{vitto_biryukov} with a \texttt{BN254} curve, \tool{} \ale{would only require} storing a single 256-bit accumulator value and, for each claim to be included in the VC, a 256-bit witness.
As the number of claims increases, our solution consumes less storage compared to SD-JWT. In Section \ref{sec:evaluation}, we comprehensively demonstrate that the reduction in storage space is about $(N - 2) \times 256$ bits, where $N$ represents the number of claims included in the VC.}

If the holder could perform runtime generation of $W$, we could further reduce the storage requirements of \tool{}. However, with the current state of cryptographic accumulators, the holder cannot generate $W$ while keeping the accumulated value at a fixed length \cite{ghosh_2016}. This limitation is fundamentally tied to trapdoor-based accumulators, as enabling the holder to generate witnesses would require sharing the private key $sk_i$, allowing them to craft valid witnesses for claims that have not been certified by any trusted party $i$. Therefore, to achieve runtime generation of constant-size $\mathcal{W}$, a trapdoor-less accumulator with fixed-length witnesses is needed, but such a construct does not currently exist.

\smallskip
\noindent \textbf{Computing Requirement.}
\tool{} does not require intensive computations, making it suitable for constrained scenarios, such as \review{hardware} wallets or IoT devices, where computing capabilities may be limited. Using the accumulator, \tool{} does not impose any overhead on the holder side, as all the operations involving direct computation over this cryptographic structure (i.e., accumulating claims and generating corresponding witnesses) are performed by the issuer. 

On the other hand, generating VPs, as seen in SD-JWT and other approaches, must \car{be} performed by the holder as they are required to prove ownership of the disclosed claims. However, in \tool{}, the holder performs minimal computation, being only responsible for constructing and signing the JWT used for presentation, which contains the pairs $(c_j, w_j)$. Thus, the most resource-intensive operation for the holder is signing the JWT with their private key $sk_h$, which is the minimum requirement also for SD-JWT.

\smallskip
\noindent {\car{\textbf{Revocation.}}} \car{Credential revocation refers to the process of invalidating a previously issued credential so that it can no longer be trusted or used by the identity holder. Revocation mechanisms apply to the entire credential and are \emph{orthogonal} to selective disclosure techniques. For example, Revocation List 2020 \cite{rl} maintains bitstrings where each bit corresponds to the status of a credential identifier, while \cite{mazzoccaevoke} leverages cryptographic accumulators to track valid VCs. If accumulators are employed to support both revocation and selective disclosure, this dual-purpose design can yield substantial memory saving at the library level, further minimizing overhead on resource-constrained devices.}

\car{Moreover, it is worth noting that verifiability relies on the issuer’s digital signature; thus, any modification (e.g., removal or update of a single claim) requires re-issuance of the credential, a common limitation of selective disclosure mechanisms. In SD-JWT, this entails re-signing the list of salted hashes. In CSD-JWT, revocation is handled by updating the accumulator and recomputing witnesses for the remaining valid claims.}
\section{Security Analysis}\label{sec:security}

This section analyzes the security guarantees of \tool{} against the threats identified in Section~\ref{sec:model}, \car{under standard cryptographic assumptions. We model adversaries as probabilistic polynomial-time (PPT) algorithms and express security properties in terms of negligible functions with respect to the security parameter $\lambda$.}

\subsection{Replay Attacks}

\car{To prevent impersonation, only the legitimate holder in possession of the private key $sk_h$ should be able to generate a valid $vp$. Without this guarantee, an adversary could replay previously captured $vp$ or exploit leaked data to impersonate the holder. \tool{} enforces replay protection by binding each presentation to the holder’s decentralized identifier $DID_h$ and requiring a fresh digital signature over a nonce $n$ provided by the verifier and included in the $vp$. This signature proves possession of $sk_h$ and prevents reuse of stale presentations. We assume that the holder’s signature scheme is existentially unforgeable under chosen message attacks (EUF-CMA) \cite{euf}, and that DIDs are securely bound to key pairs.}

\begin{definition}[\textbf{Replay Attack Resilience}]
\car{A selective disclosure scheme is resilient against \emph{replay attacks} if, for any PPT adversary $\mathcal{A}$ who does not possess the holder's private key $sk_h$, the probability that $\mathcal{A}$ can produce a valid $vp_\mathcal{A}$ accepted by a verifier is negligible in $\lambda$:
\[
\Pr[\mathsf{Verify}(vp_\mathcal{A}, pk_h) = 1] \leq \text{negl}(\lambda).
\]}

% \[
% \Pr[\mathsf{Verify}(vp, pk_h) = 1 \mid \mathcal{A} \text{ does not know % } sk_h] \leq \text{negl}(\lambda).
% \]}
\end{definition}

% \[
% \Pr[\mathsf{Sign}(vc, pk_h)=vp | \mathsf{Verify(vp, pk_h)=1}] \leq \text{negl}(\lambda)
% \]

\car{Let $\mathcal{A}$ be an adversary who obtains a credential bound to a legitimate holder $h$ via collusion or theft. Each credential includes $DID_h$, and any valid presentation must include a signature over a verifier-provided nonce $n$ using $sk_h$. Without $sk_h$, $\mathcal{A}$ cannot generate a valid signature due to the assumed EUF-CMA security of the signature scheme. Even if $h$ colludes with $\mathcal{A}$ and provides a valid $vp$, the verifier’s challenge with a fresh nonce $n$ ensures that previously issued proofs cannot be replayed. Thus, $\mathcal{A}$ cannot reuse any presentation without active cooperation from $h$.}

To further mitigate \car{replay} risks, additional defenses should be enforced at the service provider level. These include implementing multi-factor authentication \cite{mfa}, monitoring usage patterns to detect suspicious activities \cite{candid}, and using device attestation techniques to ensure that credentials are presented only from authorized devices \cite{seda}.

\subsection{Data Overcollection}

\tool{} enables selective disclosure by allowing the holder $h$ to reveal only a subset $C \subseteq \mathcal{C}$ of claims, along with their corresponding cryptographic witnesses $W$ and an accumulator value $a$ committing to the full set $\mathcal{C}$.

Identity owners should provide only minimal information to verifiers, such as service providers. Excessive \car{claim collection, regardless of the verifier's honesty, can increase privacy risks for the holder}. Malicious service providers may exploit the gathered knowledge either for financial gain or for profiling purposes \cite{fbcambridge}, \car{while even honest providers may inadvertently} expose users to data breaches \cite{saleem2020sok}.
\tool{} overcomes this challenge by enabling $h$ to selectively disclose only \car{a subset $C \subseteq \mathcal{C}$ of claims.} \car{This enables the verifier to validate} the authenticity and integrity of the shared information, without \car{accessing the full credential}. 

\begin{definition}[\textbf{\car{Soundness}}]\label{def:soundness}
\car{Let $\mathcal{C}$ be the set of claims committed to by the accumulator value $a$, and let $C \subseteq \mathcal{C}$ be the subset of claims disclosed by the holder $h$ during a credential presentation. A selective disclosure scheme satisfies \emph{soundness} if, for any PPT adversary $\mathcal{A}$, the probability that $\mathcal{A}$ can produce a tuple $(c_\mathcal{A}, w_\mathcal{A}, a)$ such that:
\begin{itemize}
  \item $c_\mathcal{A} \notin C$ (i.e., at least one disclosed claim $c_a$ is not part of the original credential),
  \item $w_\mathcal{A}$ is a valid witness for $c_\mathcal{A}$ with respect to $a$,
  \item and $\mathsf{VerifyWitness}(a, c_\mathcal{A}, w_\mathcal{A}, pk_i) = 1$,
\end{itemize}
is negligible in the security parameter $\lambda$:
\[
\Pr\left[
\begin{array}{c}
\mathsf{VerifyWitness}(a, c_\mathcal{A}, w_\mathcal{A}, pk_i) = 1 \\
\land\ c_\mathcal{A} \notin C
\end{array}
\right] \leq \text{negl}(\lambda).
\]}
\end{definition}

\begin{definition}[\textbf{\car{Information Minimality}}]
\car{A selective disclosure scheme satisfies \emph{data minimization} if, during a verification session, the verifier $v$ learns only:}
\begin{itemize}
  \item \car{the subset of claims $C \subseteq \mathcal{C}$ intentionally disclosed by the holder $h$, and}
  \item \car{auxiliary information (e.g., witnesses $W$, accumulator $a$) that is computationally independent of $\mathcal{C} \setminus C$ and its \car{cardinality} $|\mathcal{C} \setminus C|$.}
\end{itemize}
\end{definition}

\car{In \tool{},} during the presentation of a credential, the holder $h$ transmits the disclosed claims $C$, cryptographic witnesses $W$, and an accumulator value $a$ committing to the full set $\mathcal{C}$ without revealing any additional information. \car{ Crucially, the proof system guarantees that the verifier can validate the authenticity of the disclosed claims without learning anything about the undisclosed portion $\mathcal{C} \setminus C$. This is achieved through two key properties: (i) \emph{soundness}, ensuring that each disclosed claim in $C$ is verifiably part of the committed set $\mathcal{C}$, and (ii) \emph{information minimality}, meaning that $W$ and $a$ reveal no information about the content, structure, or size of $\mathcal{C} \setminus C$. As a result, \tool{} fully satisfies the data minimization definition by limiting verifier knowledge to only what is explicitly disclosed, while preserving the privacy of all other credential data. }

 \car{It is worth noting that structural metadata, such as position or number of undisclosed claims, can itself leak sensitive information, as recently highlighted in the SD-JWT IETF draft \cite{ietf-oauth-selective-disclosure-jwt-22}. To mitigate this, SD-JWT recommends adding decoy digests (i.e., mock claims) when a real claim is conditionally included. Specifically, if a claim is only present when a specific condition is met, the issuer should include a decoy digest when the condition is not met, to prevent revealing the absence of the claim.} 
 
 \car{Let us consider a scenario where revealing the claim number can be exploited to enable manipulative practices or unfair decision-making.} For instance, an individual holds a credential issued by a healthcare provider containing 20 claims related to their health condition. \car{To qualify for a health-based discount,} the holder \car{wishes} to share only \car{two claims (i.e., health center and the last check-update with a fitness center)}. \car{However,} by providing the entire credential for verification, $v$ can observe that many claims \car{are being withheld.} This observation \car{alone may} lead the verifier to speculate about the sensitive nature of the \car{undisclosed} information, even if those claims were not relevant to the service.

\tool{} is resilient by design against this class of attacks because beyond the disclosed claims $C \subseteq \mathcal{C}$, the only additional information revealed is $a$ and $W$. This information does not leak anything about the full set of claims, specifically how many of them are contained in it. \car{In contrast, SD-JWT can achieve similar privacy guarantees through the use of decoys; however, this comes at the cost of increased credential size, making \tool{} even more efficient and impactful.}

\subsection{Compromised Communication Channel} 
To prevent \car{passive eavesdropping}, secure communication among parties must be established. This can be effectively achieved by employing asymmetric encryption, utilizing the public keys stored in the DID Document of participants.
\car{Let $\mathcal{A}$ be an adversary attempting to tamper with the accumulator value $a$ or $WVC$. Any modification would require generating a valid witness for the altered data, which is infeasible without the issuer’s private key $sk_i$ under the EUF-CMA assumption. This ensures that any tampering with credential components is detectable and rejected by the verifier. \tool{} guarantees both binding and hiding properties, preserving both integrity and privacy.}

\section{Performance Evaluation}\label{sec:evaluation}
In this section, we present a comprehensive evaluation of \tool{}. 
We performed a series of experiments to assess the impact of the proposed mechanism on the primary actors within the SSI ecosystem.  Specifically, the following experiments were conducted: 

\begin{itemize}
    \item \textbf{Credential Issuance}: Evaluated the overhead for issuers to issue VCs using \tool{};
    \item \textbf{Credential Storage}: Measured the storage requirements for a credential, highlighting \tool{} feature of facilitating selective disclosure with minimal storage needs.
    \item \textbf{Generating and Verifying Verifiable Presentations}: 
    Assessed the latency and network overhead for presenting and verifying VPs. 
\end{itemize}

Experiments were conducted by comparing \tool{} to SD-JWT and run on a 14th Gen Intel(R) Core(TM) i9-14900k \asia{that features} 8 p-cores and 16 e-cores, and 48GB of RAM.
\asia{For holder-side operations, we utilized the Nordic nRF52840 \cite{nordic_nrf52840}, an RFC7228 class-2 constrained device \cite{rfc7228}. The nRF52840 is equipped with a 32-bit ARM Cortex-M4 processor, 1 MB of Flash memory, and 256 KB of RAM.}
\review{We compared \tool{} to SD-JWT as it represents the current state-of-the-art solution and remains the only fully implemented and widely recognized solution available in the field at the time of this work}. The experimental findings demonstrate that our mechanism is a valuable solution to enable selective disclosure while \ale{enhancing} holder privacy \ale{compared to SD-JWT}. For fairness, all comparisons between \tool{} and SD-JWT were conducted using the same claims within the credentials. All experiments were repeated \ale{$100$} times, and the results were averaged. \car{A more extensive performance evaluation focusing also on Merkle Tree (MT) and BBS+ mechanisms for selective disclosure is provided in Appendix~\ref{app:extensive_evaluation}.}

\begin{figure*}[!t]
    \centering
    \begin{subfigure}[t]{0.45\textwidth}
        \centering
        \includegraphics[width=\linewidth]{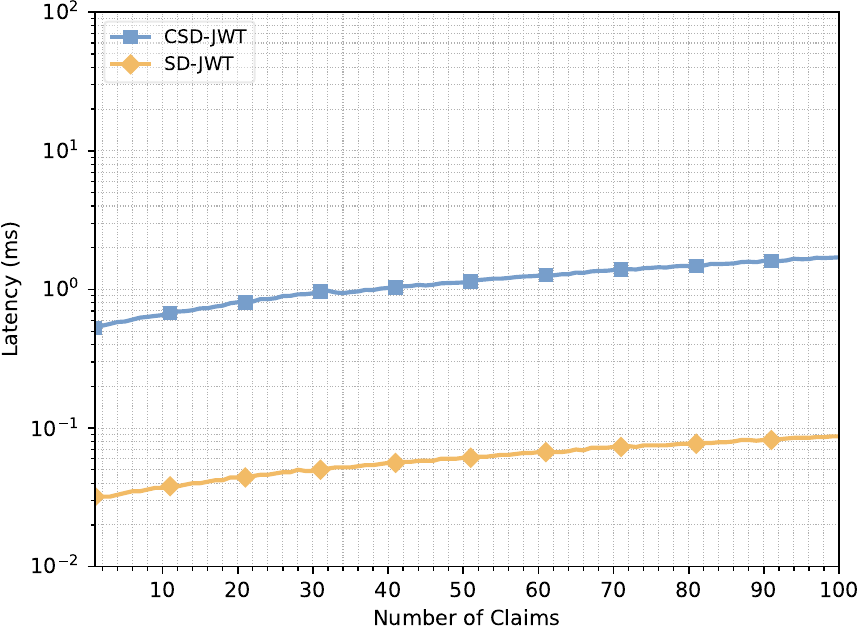}
        \caption{Issuer overhead (ms) for generating VCs.}
        \label{fig:VC_generation_time}
    \end{subfigure}
    \hfill
    \begin{subfigure}[t]{0.45\textwidth}
        \centering
        \includegraphics[width=\linewidth]{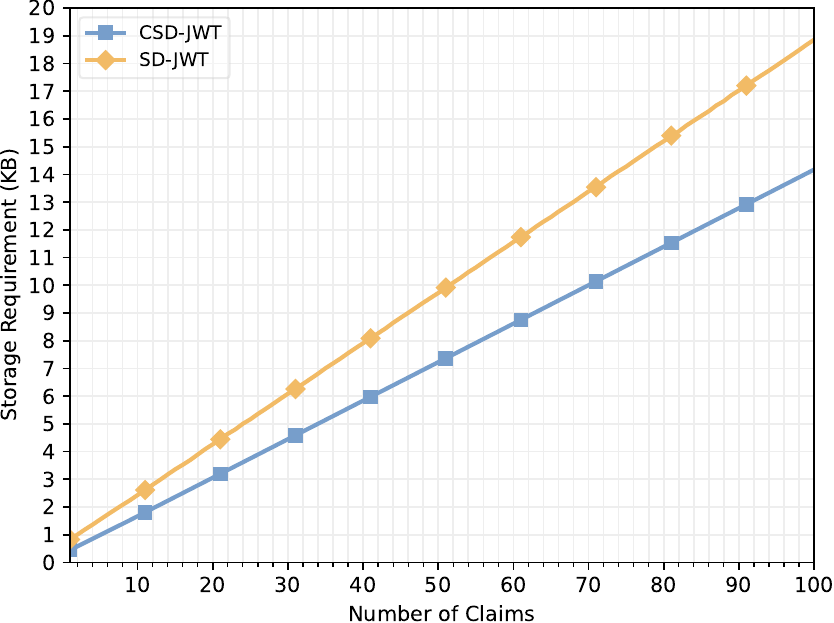}
        \caption{Storage requirements (KB) for VCs.}
        \label{fig:storage_requirement}
    \end{subfigure}
    \caption{Performance metrics for VC issuance and storage.}
    \label{fig:VC_metrics}
\end{figure*}

\subsection{Implementation Setup} 
We implemented a prototype of \tool{}\ale{\footnote{\url{https://github.com/csdjwt/csd_jwt}}  \footnote{\url{https://github.com/csdjwt/csd_jwt_artifact_evaluation}}} and evaluated its performance while delivering the core functionalities. 
 To effectively build the SD-JWT mechanism, we relied mainly on two libraries: \texttt{josekit}\footnote{\url{https://crates.io/crates/josekit}}, an OpenSSL-based library that provides ease of access to cryptographic signatures, and \texttt{serde\textunderscore json}\footnote{\url{https://crates.io/crates/serde_json}}, a library that provides APIs to serialize and deserialize Rust objects to a JSON format. For generating accumulator values and the witnesses associated with claims, we relied on \texttt{vb-accumulator}\footnote{\url{https://crates.io/crates/vb-accumulator}}, developed by DockNetwork, \asia{which} meets the requirements of \cite{vitto_biryukov}. 
\car{The underlying cryptographic accumulator is the ECC-based construction proposed in \cite{vitto_biryukov}, which supports efficient batch operations for both element accumulation and witness generation. The accumulator value is represented as a 256-bit elliptic curve point over the BN254 curve. To the best of our knowledge, this choice offers an optimal tradeoff among security guarantees, compact accumulator and witness sizes, and computational efficiency, making it particularly suitable for constrained environments.}
\asia{To evaluate the feasibility of \tool{} and SD-JWT on constrained holders, we developed a framework\ale{\footnote{\url{https://github.com/csdjwt/csd_jwt_constrained}}} in C using the nRF Connect SDK and Parson\footnote{\url{https://github.com/kgabis/parson}}, a lightweight library for managing JSON data structures.}
Our credentials can include a variable number of claims\review{, each formatted as \texttt{claim\_key:claim\_value}. For our evaluation, we consider VCs with up to 100 claims, reflecting variations across application domains \cite{EU_Digital_Identity_Wallet_Use_Cases}. For instance, a VC representing an identification document typically contains around 15 claims. In contrast, in healthcare scenarios, the number of claims can vary significantly, ranging from approximately 10 claims, as in the case of the healthcare passport established by many countries for COVID-19 \cite{covid19}, to over 100 in detailed medical reports \cite{THOMASON2021165}.} Finally, for transparency and reproducibility, we open-sourced the proposed implementations. 

\subsection{Credential Issuance} 
To simulate credential issuance, we evaluated the issuer overhead associated with issuing a valid VC. 
Specifically, we consider the latency to accumulate claims in the accumulator, embed the accumulator value within the VC, and generate witnesses. For SD-JWT, we measured its latency to generate salts with a pseudo-random number generator, compute hashes for all the claims to certify their validity, and produce an ECDSA signature.

Figure \ref{fig:VC_generation_time} compares the issuance latencies, expressed on a logarithmic scale, of the considered methods. \asia{SD-JWT achieves slightly better performance, mainly due to the CPU instructions used by the two mechanisms. Indeed, despite the great improvements in the instruction set of processors regarding finite field operations, hashing algorithms remarkably outperform them. However, it is worth noting that \tool{} only requires a few milliseconds even for credentials including 100 claims. Moreover, this does not represent a serious concern since, in real-world scenarios, issuers typically have powerful computing resources to manage the issuance process, significantly reducing these latencies.}

\subsection{Credential Storage} 
One of the key properties of \tool{} is its ability to minimize the amount of disclosed data, making it resilient against inference attacks and reducing the size of the VC. This compactness is especially crucial in scenarios where holders have restricted storage capabilities, such as hardware wallets or IoT devices.
We conducted a series of experiments varying the number of claims within the VC, from 1 to 100. For each generated credential, we evaluated the overall storage requirements for the two compared methods. \car{For our mechanism, this encompasses the storage of the accumulator value and the WVC, which contains the pairs of claims and witnesses.} The storage requirements for SD-JWT comprise the signed list of hashes and the Salt-Value Container (SVC), which includes the pairs of claims and salts. 

As shown in Figure \ref{fig:storage_requirement}, \tool{} remarkably outperforms SD-JWT even with a low number of claims in the VC. Using identical claims for both mechanisms, the primary difference lies in the size of the credentials and the storage of witnesses versus salts. The storage requirement for our method is much lower because \tool{} only contains the accumulator value, which is 256 bits. In contrast, SD-JWT comprises the hashes of the $N$ claims, i.e., $N \times$ 256 bits, making it dependent on the number of attributes. On the other hand, salts used in SD-JWT only require $N \times$ 128 bits, against the $N \times$ 256 bits for witnesses. 
Therefore, SD-JWT \asia{brings} an overhead of ($N - 2) \times 256$ bits for storing a VC with the same claims. As highlighted in the figure, \tool{} saves a significantly larger portion of memory while increasing the number of claims, achieving up to $46\%$ storage reduction. These results demonstrate that \tool{} is a valuable solution for selectively disclosing credentials in constrained environments. Moreover, this saving is even more impactful as each holder is expected to maintain multiple credentials \cite{gartner2024digitalidentity, dayaratne2024ssi4iot}.

\begin{figure*}[!t]
\centering
\begin{subfigure}{0.30\textwidth}
    \centering
    \includegraphics[height=0.18\textheight]{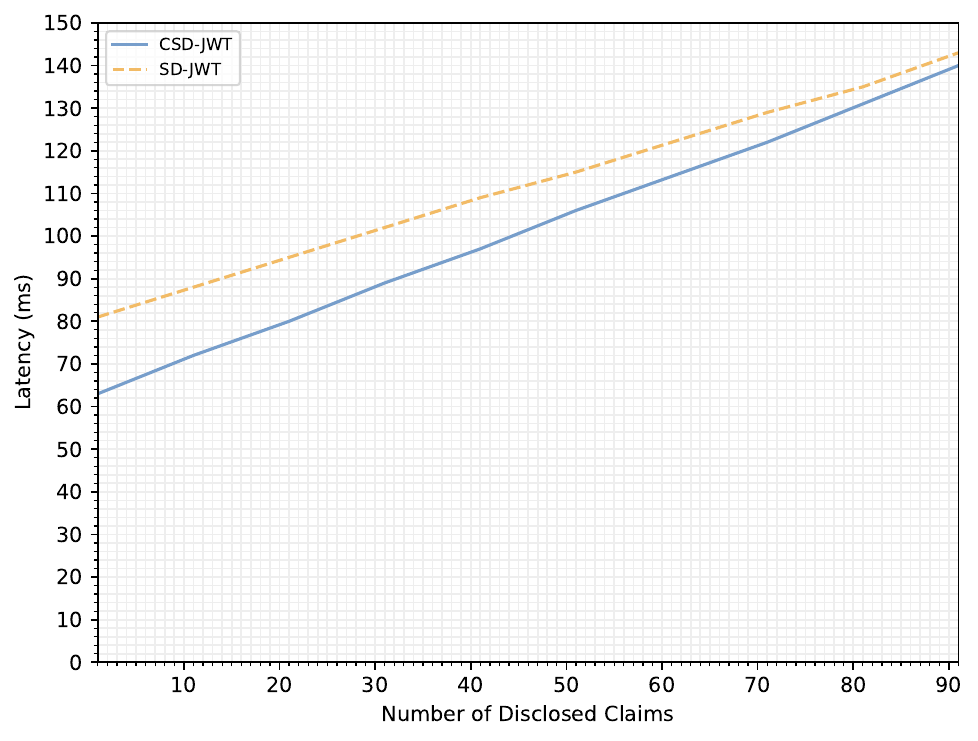}
    \caption{VP generation (ms) for a credential with 100 claims on a constrained device.}
    \label{vp_latency_generation}
\end{subfigure}
\hfill
\begin{subfigure}{0.30\textwidth}
    \centering
    \includegraphics[height=0.18\textheight]{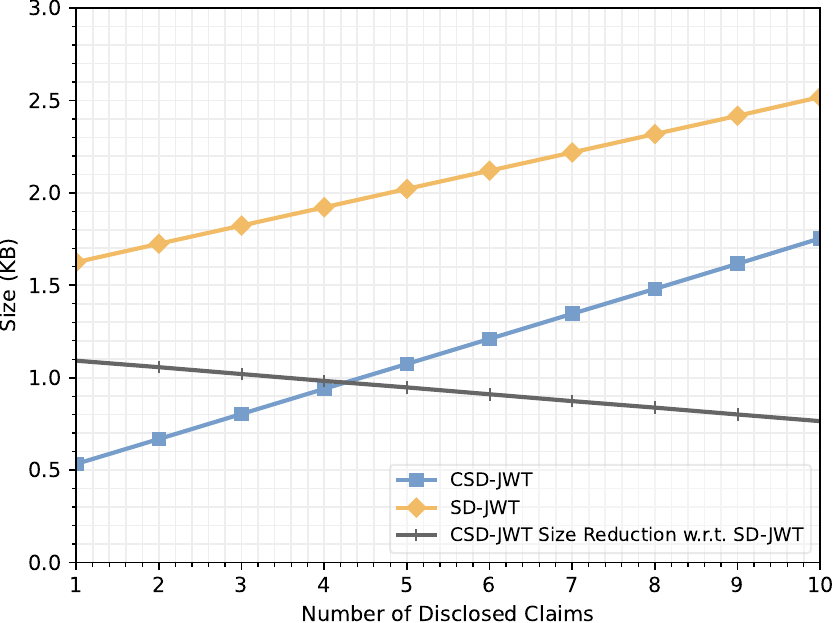}
    \caption{VP sizes (KB) based on a credential with 10 claims.}
    \label{10_claims_vp_size}
\end{subfigure}
\hfill
\begin{subfigure}{0.30\textwidth}
    \centering
    \includegraphics[height=0.18\textheight]{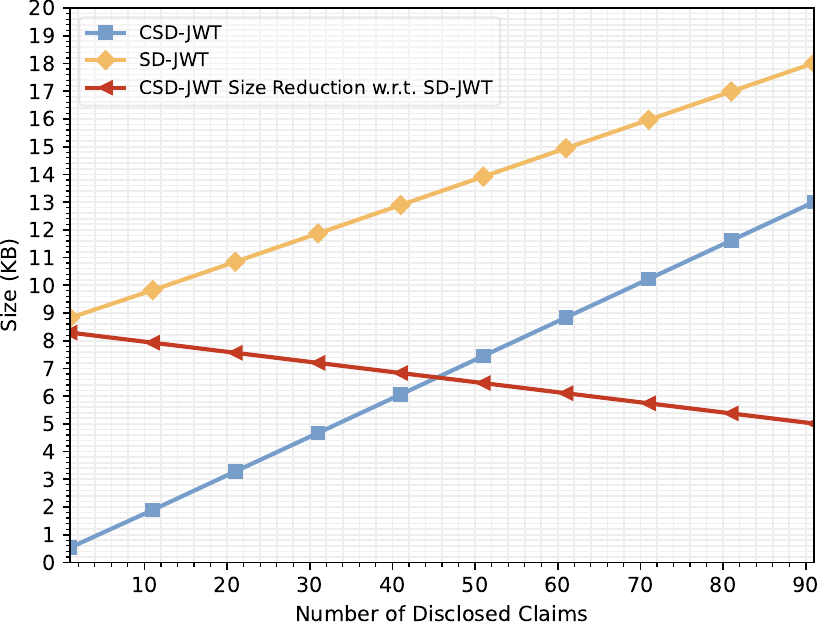}
    \caption{VP sizes (KB), based on a credential with 100 claims.}
    \label{100_claims_vp_size}
\end{subfigure}
\caption{Comparison of \tool{} and SD-JWT for VP generation (a) and size (b),(c) across different scenarios.}
\label{combined_vp_size}
\end{figure*}

\subsection{Generating and Verifying Verifiable Presentations}
The compactness of \tool{} reduces the storage requirements for holding credentials and decreases the size of the VPs shared with the verifier. These experiments aim to \review{evaluate} the \asia{generation time and }size of VPs, \review{produced from VCs having a different number of claims}, while varying the percentage \review{of disclosed ones}. Specifically, \review{we generated 10 VCs each having a number of claims equal to a multiple of 10, starting from 10} up to 100 and, for each \review{of them}, we varied the percentage of disclosed claims from $0\%+1$ \review{of the total} to $90\%+1$.

\smallskip
\noindent \textbf{\asia{Generation of Verifiable Presentations.}} \asia{We conducted experiments to evaluate the latency of an RFC7228 class-2 constrained device in generating VPs using both selective disclosure techniques. \tool{} and SD-JWT perform similar operations to generate VPs, as both involve selecting claims to disclose and signing them. This similarity is reflected in the experimental results shown in Figure \ref{vp_latency_generation}, where \tool{} achieves slightly better performance due to the smaller size of the processed data. As anticipated, memory usage metrics for the firmware also show no significant differences, with RAM consumption remaining around 18\%.}

\begin{figure}[!t]
\centering
\includegraphics[width=0.90\columnwidth]{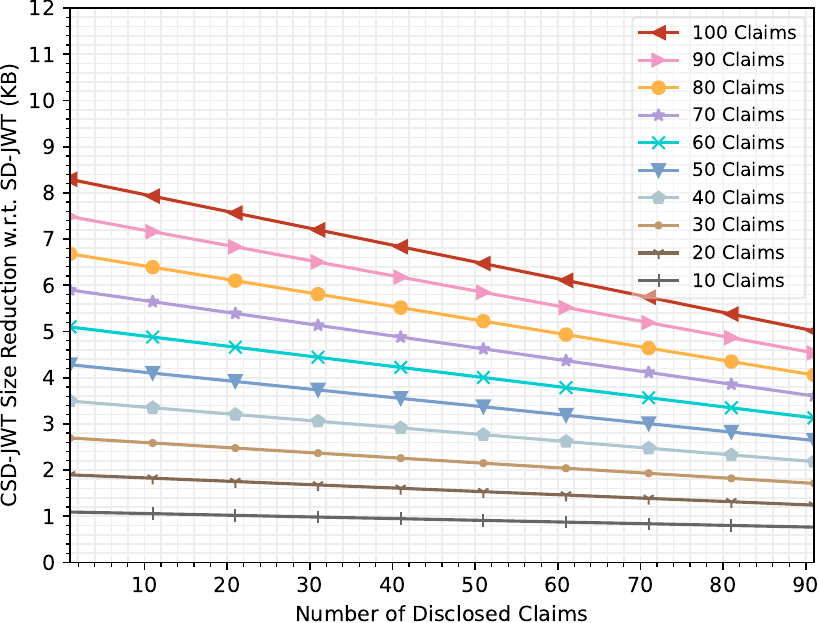}
\caption{\asia{VP size reduction of \tool{}.}}
\label{difference_in_vp_size}
\end{figure}

\smallskip
\noindent \textbf{Size of Verifiable Presentations.} Figures \ref{10_claims_vp_size} and \ref{100_claims_vp_size} report the results for VCs containing 10 and 100 claims, respectively. As clearly shown, \tool{} consistently generates smaller VPs than SD-JWT regardless of the number of disclosed claims. In Figure \ref{10_claims_vp_size}, when the holder \review {discloses a single} claim included in a VC with 10 claims, \tool{} achieves a $67\%$ reduction in the VP size, which decreases to $30\%$ when \review{all the claims included in the VC are disclosed}. The savings in the VP size are even more significant \review{when considering} VCs \review{including} a larger number of claims. Figure \ref{100_claims_vp_size} demonstrates that with \tool{}, a VP disclosing 1 claim out of 100 is \asia{$93\%$} smaller \review{in size} than \review{ the same VP produced with} SD-JWT. Even when all claims are being disclosed, \tool{} still provides a $27\%$ size reduction. Figure \ref{difference_in_vp_size} illustrates the reduction in VP size by varying the rounded percentage of disclosed claims. These results highlight that \tool{} is an effective method to generate VPs \review{with reduced size and network overhead, a feature that is} particularly important in scenarios where holders\review{' devices} have limited \review{memory and} network capabilities.

\begin{figure}[!t]
\centering
\includegraphics[width=0.90\columnwidth]{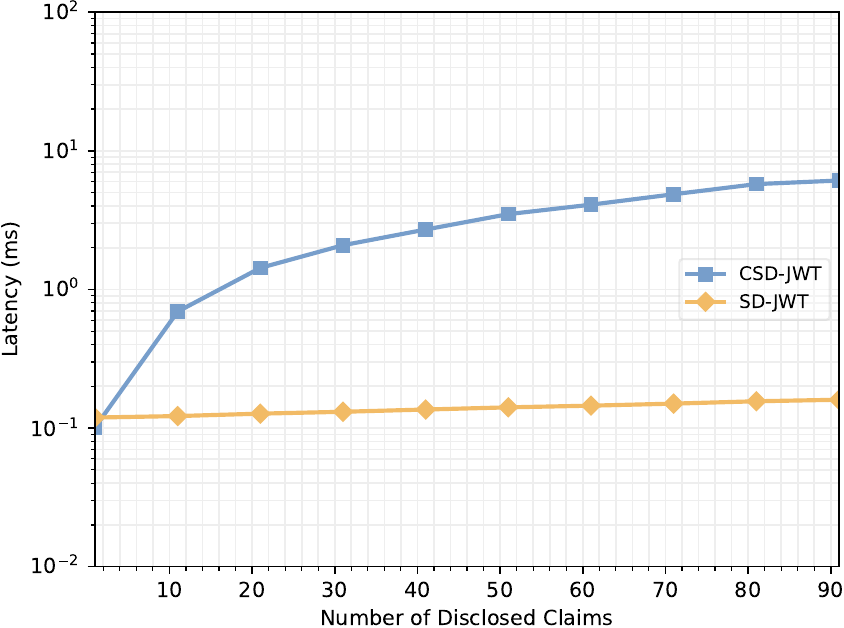}
\caption{\ale{VP verification latency for a VC storing 100 claims.}}
\label{verification_latency}
\end{figure}

\smallskip
\noindent \textbf{Verification of Verifiable Presentations.} Finally, we evaluated the time required for a verifier to assess the authenticity of the presentations. Figure \ref{verification_latency} shows the verification latencies for claims of different sizes using both SD-JWT and \tool{}. The latency for verifying a VP generated with SD-JWT is negligible \asia{(up to a few $\mu$s)} since the procedure involves the verification of a digital signature, a handful of hashing operations, and a comparison with a list of hashes. In \tool{}, verification time is affected by the underlying cryptographic accumulator. However, it remains on the order of a few milliseconds, meeting the response requirement of the Resource Animation Idle Load (RAIL) model proposed by Google\footnote{\url{https://web.dev/articles/rail}}. \asia{As a result, the overhead introduced by \tool{} is minimal and comparable to that of SD-JWT. Furthermore, verification is performed by verifiers, which are entities typically equipped with more powerful computational resources, making the impact of such overhead even less relevant.}

\subsection{Discussion of Results}
This section provides further consideration of the conducted experiments. It is worth noting that employing different implementations could lead to improved performance. 

\smallskip
\noindent \textbf{Credential Issuance.} The time required to issue credentials based on CSD-JWT is higher than that for SD-JWT credentials, due to the increased complexity of the operations involved. In particular, the most demanding operation is the witness generation that, as demonstrated in \cite{camacho2010impossibility}, for $w$ witnesses demands at least $\Omega(w)$. However, the issuance time is a few milliseconds and does not represent a concern, as trusted authorities acting as issuers may have more powerful computing resources than those used for our experiments. Furthermore, it is worth noting that as these operations are always done on the issuer side, they do not affect at all the computing capabilities of the holder. 

\smallskip
\noindent \textbf{Credential Storage.} In \tool{}, the storage requirement for the \car{accumulator value} \asia{remains} constant, regardless of the number of claims included. \asia{This contrasts with} SD-JWT, where \ale{the signed list} contain a \asia{salted} hash for each claim. \asia{Although} salts \asia{are} smaller in size compared to the witnesses, \asia{\tool{} offers an \car{overall} size reduction ratio of $27-93\%$ for VCs with 100 claims, with even greater savings for credentials containing more fields.} 
\asia{In constrained environments, resources are shared across different functions. For instance, it is essential that the memory securely stores VCs and firmware, making even KB-level reductions valuable. Additionally, minimizing resource usage improves energy efficiency, a critical consideration in resource-limited settings. Finally, while our evaluation focuses on the overhead reduction for a single VC, identity owners are projected to use multiple VCs by 2026 \cite{gartner2024digitalidentity}, making the reductions offered by \tool{} significantly more remarkable.}

\smallskip
\noindent \textbf{Verifiable Presentation.} The generation of VPs must be performed directly by holders. This is a logical consequence of the fact that they have to prove ownership over the disclosed information through a secret key. However, creating a VP does not introduce a particular overhead. Similar to SD-JWT, the holder is only required to create and sign a JWT containing the claims to be disclosed. \asia{This makes both \tool{} and SD-JWT practical on constrained devices.}

On the other hand, we observed \asia{remarkable} differences in \ale{size of the respective VPs}. In SD-JWT, the \asia{credential} shared during the presentation comprises the hashes of all the claims; thus, the size depends on the number of \asia{data} included. \car{In contrast, \tool{} maintains a constant-size accumulator value regardless of the number of claims, ensuring compactness. The only component that grows linearly with the number of disclosed claims is the set of witnesses required for verification.} This minimizes the information shared, significantly reducing the network overhead and offering higher privacy.

\section{Related Work}\label{sec:related}

\asia{This section reviews the primary selective disclosure mechanisms for VCs, categorizing them based on the underlying cryptographic techniques used to disclose claims. The categories include atomic credentials, hashed values,  signature schemes, and Zero-Knowledge Proofs (ZKPs)} \cite{ETSI}.

\subsection{Atomic Credentials}

Atomic credentials, or monoclaim credentials, \asia{represents the simplest} selective disclosure technique. \asia{A multiclaim credential is partitioned} into a subset of monoclaim credentials. Instead of having a single VC with $N$ claims, the holder will have $N$ VCs, each containing a single claim. Although this mechanism is compatible with existing solutions that do not yet support selective disclosure, it does not scale with credentials holding many data fields. 
\asia{Managing $N$ distinct credentials increases storage requirements, as each credential must include its metadata, cryptographic signatures, and auxiliary data. This also affects verification, as verifiers must individually validate each credential, raising computational and time costs. Furthermore, this approach incurs remarkable network overhead, as transmitting multiple credentials to disclose claims negatively impacts communication efficiency in bandwidth-constrained environments.} \asia{These challenges are addressed by more efficient mechanisms that do not need to generate multiple credentials to selectively disclose claims.}

\subsection{Hashed Values}
\asia{In} hash-based methods, the claims in a VC are replaced with their salted hashes. 
SD-JWT \cite{SD-JWT} is the state-of-the-art solution of JWT for selective disclosure, where each claim in the JWT payload is combined with a unique salt, \asia{hashed, and included in the credentials}, ensuring claims are not directly exposed. \asia{The holder is given an SVC, containing the plaintext claims and their corresponding salts. To disclose claims, the identity owner provides the verifier with the SD-JWT, the plain text claims they wish to disclose, and the corresponding salts. The verifier hashes each claim with its salt and checks if they match the pre-images of the hashed values in the SD-JWT.} The main downside of this method is the credential size, which \asia{grows} with the number of claims, impacting storage and transmission efficiency. Moreover, including hashes for all claims in the SD-JWT can make the system susceptible to inference attacks \asia{based on the credential's structure. Similar to SD-JWT, De Salve et al. \cite{desalve} propose using Hash-based Message Authentication Codes (HMACs) \cite{turner2008keyed} to replace plaintext claims in VCs.
The holder receives plain text claims and keys used to generate the HMACs. During the presentation, the holder discloses selected claims and keys, allowing the verifier to regenerate HMAC and validate HMAC. However, this approach shares the same limitations as SD-JWT. These drawbacks are addressed by \tool{}, which prevents leakage of additional information (e.g., claim number) and substitutes the list of salted hashes with an accumulator value, whose size is constant} regardless of the number of included claims.

\asia{Merkle Trees \cite{merkle_tree_definition} can also be used to create valid proofs for disclosed claims \cite{sd_merkle_tree}. Each leaf node represents a salted claim's hash, and parent nodes are obtained by hashing their children. Thus, knowing the root hash, a verifier can confirm a claim's inclusion using a proof consisting of sibling hashes along the path to the root. This method ensures fast verification while keeping claims hidden. However, unlike \tool{}, it results in variable proof lengths as the tree size grows with the number of claims in the credential. For binary trees, the path length is $log_2(N)$, where $N$ is the number of claims, potentially revealing a range for the VC's field count.}

\subsection{Signature Schemes}

\asia{Signature schemes like Camenisch-Lysyanskaya (CL) signature \cite{cl_sig_1} are widely used in Anonymous Credentials (ACs) \cite{kakvi2023sok}, which can be seen as a special type of VCs, to ensure privacy guarantees such as anonymity or unlinkability. These schemes enable the holder to reveal specific claims while preserving anonymity. However, the applicability of ACs is limited being the authenticity of an identity or claim a primary requirement in most real-world scenarios \cite{vc_use_case}.} 

\asia{Historically, the foundational work was done by the CL signature \cite{cl_sig_1}, which relies on the hard RSA assumption. Each message $m_i$ is mapped onto a finite field using modular exponentiation on random values $a_i, b, s, c$ to generate the signature. Given the public key $pk = (n, a_1, \dots, a_n, b, c)$ and the signature $(v, e, s)$ the verification is performed as follows:}

\begin{equation}
v^e \equiv a_1^{m_1}\dots a_n^{m_n} b^s c \mod n    
\end{equation}

\asia{For selective disclosure, the public key reveals $a_i^{m_i}$ rather than $a_i$. Several other schemes have been proposed \cite{bbs, bbs_plus, cl_sig_2, idemix} with most of them using bilinear maps, allowing holders to derive valid signatures for subsets of signed messages. This prevents issuer-verifier collusion but introduces high computational and storage overhead, especially on the holder side, making them less suitable for constrained environments. Such concerns do not affect \tool{}, as the holder is only required to select and sign the VP, without performing complex cryptographic operations.}

\subsection{Zero-Knowledge Proofs}

\asia{ZKPs \cite{first_zkp} enable a prover to convince a verifier of the truth of a statement while concealing any information beyond its validity. Thus, they are often leveraged alongside signature schemes to prove a possession claim without revealing the actual information. Specifically, research has focused on Zero-Knowledge Succinct Non-Interactive Arguments of Knowledge (zk-SNARKs) \cite{first_zk_snarks}, non-interactive methods to produce ZKPs, where the prover produces proofs by encoding problems as circuits.} \asia{Schanzenbach et al. \cite{zklaims} proposed ZKlaim, \asia{which} uses zk-SNARKs with Groth's scheme \cite{groth_2016} to produce succinct proofs from an issuer's constraint system. While verifying proofs is efficient, their generation is time-consuming and requires significant storage. Lee et al. \cite{lee_zk_snarks} proposed a more efficient commit-and-prove solution, though time and space efficiency remain limited compared to hash-based methods and \tool{}, where VP generation consists of selecting and \car{signing} claims to disclose.}

\section{Conclusion}\label{sec:conclusion}

\review{
In this paper, we proposed \tool{}, a novel selective disclosure mechanism that leverages a cryptographic accumulator \asia{to minimize the size of VCs, enabling the identity owner to reveal only strictly necessary information.} We implemented \tool{} as an open-source solution} and conducted a comprehensive evaluation against the current state-of-the-art, SD-JWT. Experimental results demonstrate that \tool{} generates compact VCs and VPs that ensure significant storage savings and minimize network overhead. These enhanced features make it particularly suitable for hardware wallets and IoT devices that usually have limited storage, computing, and network capabilities.

\section*{Acknowledgment}~\label{ack} 
This work was partially supported by the SERICS project (PE00000014), funded under the MUR National Recovery and Resilience Plan program funded by the European Union - NextGenerationEU, and the U.S. National Science Foundation through the Intergovernmental Personnel Act Independent Research \& Development Program.
The views expressed are those of the authors and do not necessarily reflect the views of the funding agencies.

\bibliographystyle{IEEEtran}
\bibliography{main} 

\ale{
\appendices
\section{Unlinkability}\label{app:unlinkability}
In the context of VCs, unlinkability refers to the inability of an adversary to associate multiple credential presentations with the same identity owner \cite{baum2024cryptographers}. This property is crucial for preserving user privacy and can be analyzed under three distinct threat models: unlinkability, unobservability, and untraceability. In the following definition,  \( \lambda \) is the security parameter and \( negl(\lambda) \) denotes a negligible probability in \( \lambda \).

\begin{definition}[\textbf{Unlinkability}]\label{def:unlinkability}
\ale{A selective disclosure mechanism satisfies unlinkability if, for a $vc$ possessed by a holder $h$, the presentation $vp_1$ generated for a verifier $v_1$ and the presentation $vp_2$ generated for a second verifier $v_2$, cannot be used by the colluding verifiers to determine that $h$ was the author of both. Formally, unlinkability holds if:
\[
\Pr[\mathcal{L}_{v_1}(vp_1, h) = 1)] \approx  \Pr[\mathcal{L}_{v_2}(vp_2, h) = 1] \leq 
\text{negl}(\lambda),
\]
\noindent where \( \mathcal{L}_{v} \) is a function executed by the verifier to attest if the holder of the credential is $h$. The function outputs 1 if $h$ is the holder that produced the verifiable presentation, 0 otherwise.}
\end{definition}

\begin{definition}[\textbf{Unobservability}]\label{def:unobservability}
A credential presentation mechanism satisfies the unobservability property if the issuer $i$ cannot determine when, where, and to which verifier \( v \) a credential is presented. Formally, unobservability holds if:
\[
\Pr[\mathcal{O}_{i}(vp, v, t) = 1] \leq \text{negl}(\lambda)
\]
\noindent where $t$ is the time of presentation, and $\mathcal{O}_i$ is a function executed by the issuer attempting to infer whether a credential it issued was presented at time $t$ to verifier $v$. The function outputs $1$ if the issuer successfully identifies the presentation event, and $0$ otherwise. 

\end{definition}

\begin{definition}[\textbf{Untraceability}]\label{def:untraceability}
A selective disclosure mechanism satisfies \emph{untraceability} if colluding issuers and verifiers cannot trace a user's credential usage across multiple interactions. Let \( vp_1, vp_2, \dots, vp_n \) be presentations derived from the same credential. Formally, untraceability holds if:
\[
\Pr[\mathcal{T}_{i,v}(vp_1, \dots, vp_n) = 1] \leq \text{negl}(\lambda),
\]
\noindent where \( \mathcal{T}_{i,v} \) is a tracing function jointly executed by colluding verifiers and issuers attempting to link multiple presentations. The function outputs $1$ if at least one colluding issuer and verifier can successfully identify the holder $h$, and $0$ otherwise.
\end{definition}

\begin{figure*}[!t]
    \centering
    \begin{subfigure}[t]{0.42\textwidth}
        \centering
        \includegraphics[width=\linewidth]{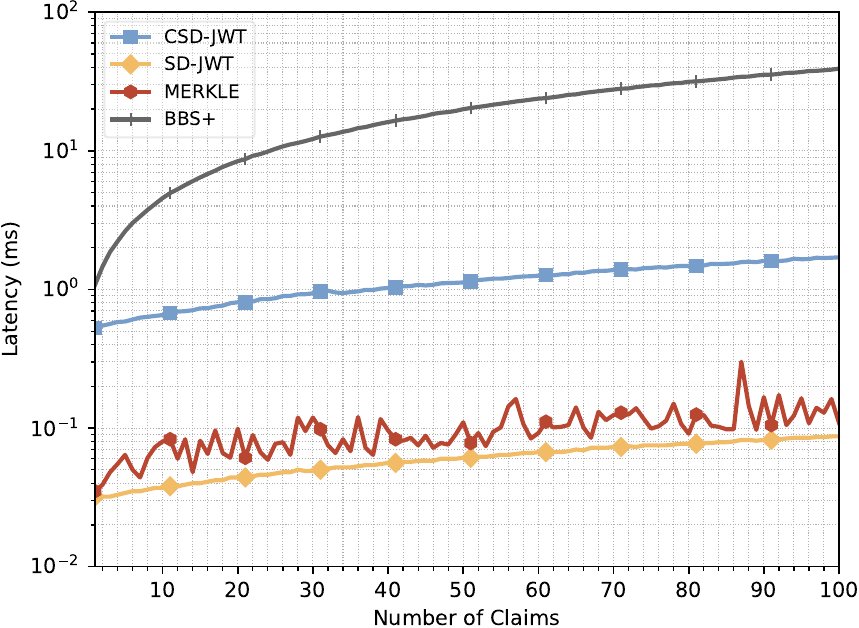}
        \caption{Issuer overhead (ms) for generating VCs.}
        \label{app:VC_generation_time}
    \end{subfigure}
    \hfill
    \begin{subfigure}[t]{0.42\textwidth}
        \centering
        \includegraphics[width=\linewidth]{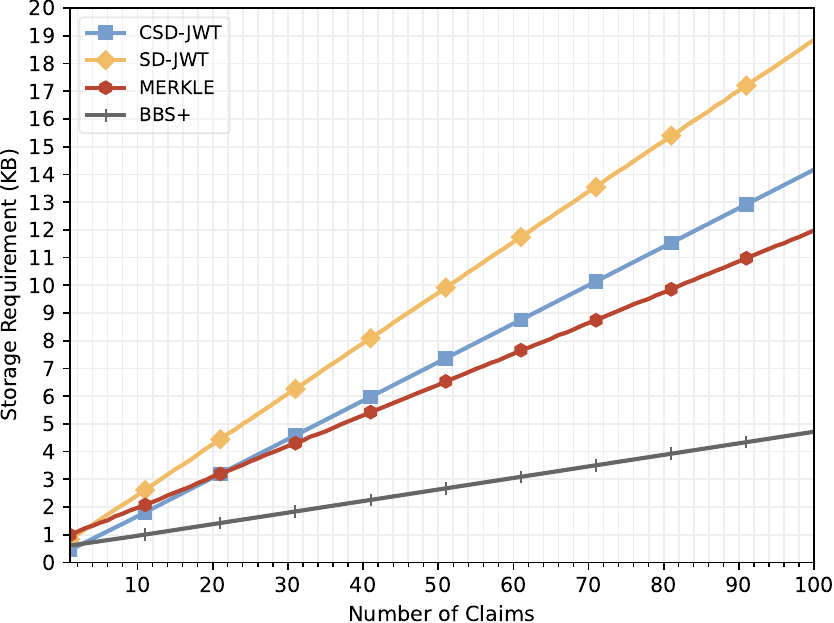}
        \caption{Storage requirements (KB) for VCs.}
        \label{app:vc_storage_requirement}
    \end{subfigure}
    \caption{Comparison of CSD-JWT, SD-JWT, MTs, and BBS+ for VC issuance and storage.}
    \label{fig:VC_metrics}
\end{figure*}

\smallskip
\noindent \textbf{Security Analysis.}
For selective disclosure mechanisms used in SSI, \emph{unobservability} is typically not a primary concern. This is because interactions between the holder and verifier are intentionally designed to avoid "phoning home" to the issuer, thereby preventing the issuer from observing credential usage. Instead, the more critical privacy properties for identity owners, whether human users or devices, are \emph{unlinkability} and \emph{untraceability}. These properties address scenarios in which entities may collect static metadata during credential presentations, store it, and subsequently collude (either across roles or with external parties) to infer sensitive information about the holder, such as their location, behavioral patterns, or activity timestamps.

Modern cryptographic constructions, such as BBS+ signatures~\cite{au2006constant}, leverage ZKPs to achieve unlinkability by ensuring that no static or identifying data is transmitted during the presentation of VPs. In contrast, mechanisms such as \tool{}, SD-JWT, and MT-based approaches are unable to guarantee unlinkability and untraceability, due to inherent limitations in their underlying designs. These approaches rely on static components embedded within the VP, such as accumulator values, witnesses, hashed attributes, salts, or Merkle paths, which may be reused or correlated across multiple interactions. As a result, they introduce linkability risks that can enable adversaries to track credential usage and compromise holder privacy.

\smallskip
\noindent \textbf{ZKP and CSD-JWT.}
To prevent tracking by colluding entities, \tool{} would need to incorporate ZKPs for both the accumulator value and the witnesses corresponding to the disclosed claims. While modern approaches, such as the one proposed in \cite{vitto_biryukov}, already support NIZKPs for concealing witnesses, they do not offer mechanisms for hiding the accumulator value. This limitation likely stems from the fact that cryptographic accumulators are traditionally used in contexts where the accumulator must remain publicly verifiable.

To the best of our knowledge, \tool{} represents the first instance where privacy-preserving selective disclosure would require the accumulator value itself to remain hidden. If future cryptographic constructions were extended to support ZKPs for both the accumulator and its associated witnesses, \tool{} could be flexibly adapted to suit different privacy and performance requirements. For scenarios prioritizing computational efficiency and energy conservation, such as those involving IoT devices, the current WVC-based approach remains preferable. Conversely, in contexts where privacy is paramount, integrating ZKPs would enable \tool{} to achieve unlinkability, significantly enhancing its privacy guarantees.
Notably, enabling unlinkability through this method would require the credential holder to generate cryptographic constructs derived from the original accumulator, introducing substantial computational overhead. As a result, while technically feasible, such an extension may be impractical for resource-constrained environments.

\begin{figure*}[!t]
    \centering
    \begin{subfigure}{0.42\textwidth}
        \centering
        \includegraphics[width=\linewidth]{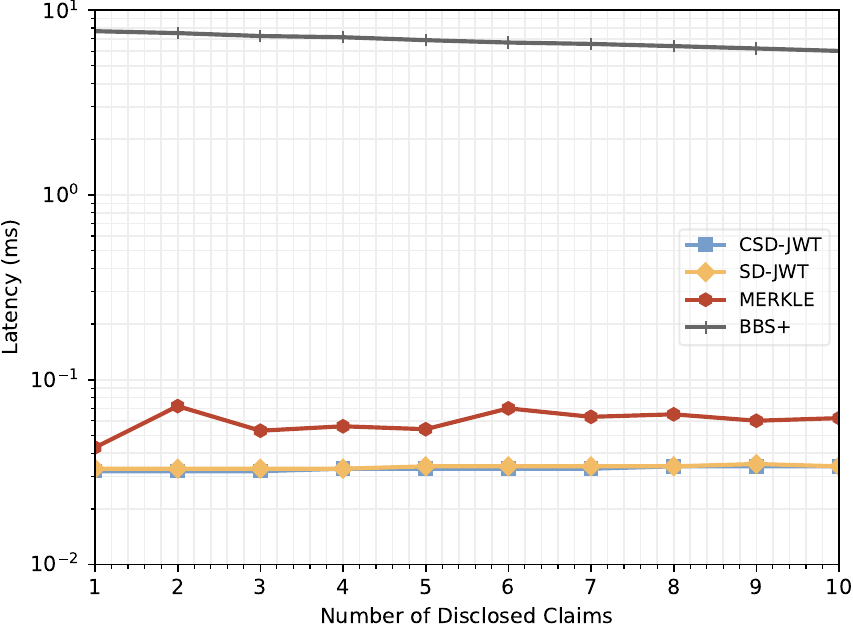}
        \caption{VP generation latency (ms) for credential with $10$ claims.}
        \label{app:10_claims_vp_issuance_latency}
    \end{subfigure}
    \hfill
    \begin{subfigure}{0.42\textwidth}
        \centering
        \includegraphics[width=\linewidth]{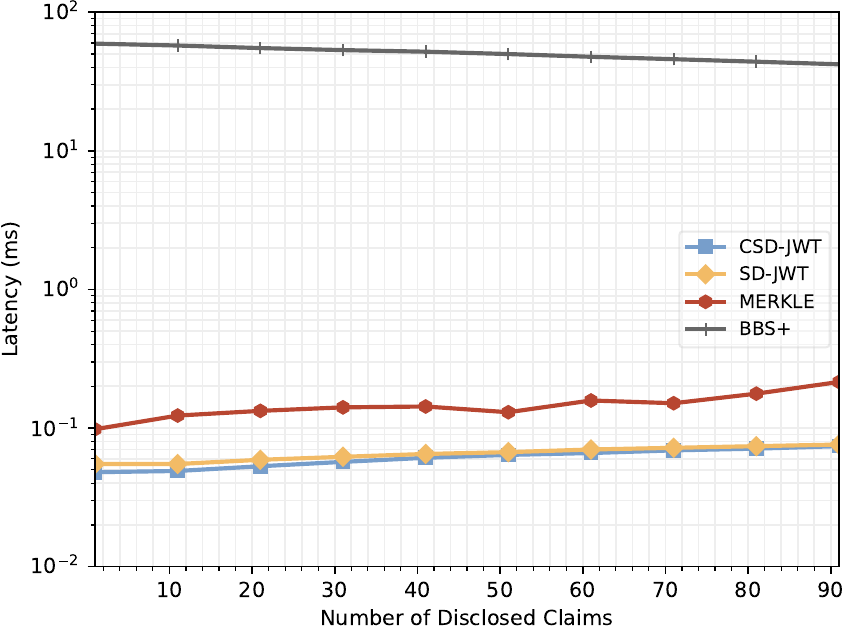}
        \caption{VP generation latency (ms) for credential with $100$ claims.}
        \label{app:100_claims_vp_issuance_latency}
    \end{subfigure}
    \caption{Comparison of \tool{}, SD-JWT, MTs, and BBS+ for VP generation across different scenarios.}
    \label{app:combined_vp_issuance_latency}
\end{figure*}

\section{Extended Performance Evaluation}\label{app:extensive_evaluation}

This section extends the performance evaluation to two additional baselines. 
Specifically, we consider the BBS+ signature scheme \cite{au2006constant}, which allows signatures on blocks of messages while enabling the proof of the validity of just a disclosed subset of them. BBS+ leverages Non-Interactive ZKPs for concealing the rest of the messages, while granting the ability to provide a Zero-Knowledge Proof of Knowledge of the signature, resulting in a fully unlinkable signature scheme that supports selective disclosure. The second mechanism leverages MTs to provide Selective Disclosure for JSON Web Proofs \cite{sd_merkle_tree}. Similar to SD-JWT, it leverages salted hashes to conceal claims. However, unlike this method, it requires building an MT stemming from the salted hashes and digitally signing its root. To disclose claims with this mechanism, the holder shall provide the path from the root down to the disclosed claims and their respective salts.

The experiments were conducted using the same setup described in Section~\ref{sec:evaluation}. For the BBS+ selective disclosure implementation, we used the Rust-based \texttt{zkryptium}\footnote{\url{https://crates.io/crates/zkryptium}} library. MT constructions were performed using the \texttt{rs\_merkle}\footnote{\url{https://crates.io/crates/rs\_merkle}} library, also implemented in Rust.

\subsection{Credential Issuance} 

Figure~\ref{app:VC_generation_time} presents a comparison of issuance latencies across the evaluated methods. SD-JWT consistently delivers the best performance, primarily due to its operational simplicity and the benefit of hardware-accelerated cryptographic primitives. It is closely followed by MT-based selective disclosure, which relies on similar primitives but incurs additional overhead from constructing the MT, which is a more complex data structure for representing claims. In contrast, \tool{} and BBS+ exhibit higher issuance latencies, attributed to their reliance on more computationally intensive cryptographic operations, such as pairing-based signatures and ZKPs. Despite these differences, all evaluated mechanisms demonstrate issuance times well below the threshold of hundreds of milliseconds, making them negligible in practical scenarios.
We argue that credential issuers typically operate on sufficiently powerful hardware to handle these operations efficiently. Moreover, given the expected issuance rate of VCs in real-world deployments, none of the considered mechanisms poses a performance bottleneck.

\subsection{Credential Storage} 
The second metric produced in this benchmark is the space required to store a VC for each of the evaluated mechanisms. As shown in Figure \ref{app:vc_storage_requirement}, all the VCs naturally scale linearly with the number of claims. In this case, though, BBS+ clearly stands out as the best achieved performance due to the nature of the construct itself: all that is required to be stored in the credential is the signature itself. Among the remaining mechanisms \tool{} remains the best solution for VCs with up to $20$ claims, while being outperformed by VCs implementing MTs afterwards, with the only consistency being SD-JWT as the worst performing among the ones evaluated.

\begin{figure*}[!t]
\centering
\begin{subfigure}{0.42\textwidth}
    \centering
    \includegraphics[width=\linewidth]{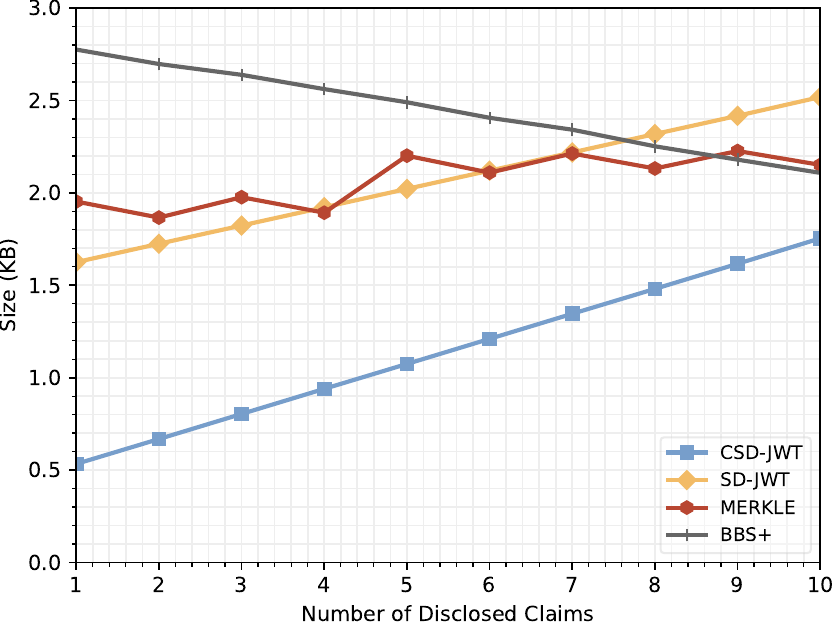}
    \caption{VP sizes (KB) based on a credential with $10$ claims.}
    \label{app:10_claims_vp_size}
\end{subfigure}
\hfill
\begin{subfigure}{0.42\textwidth}
    \centering
    \includegraphics[width=\linewidth]{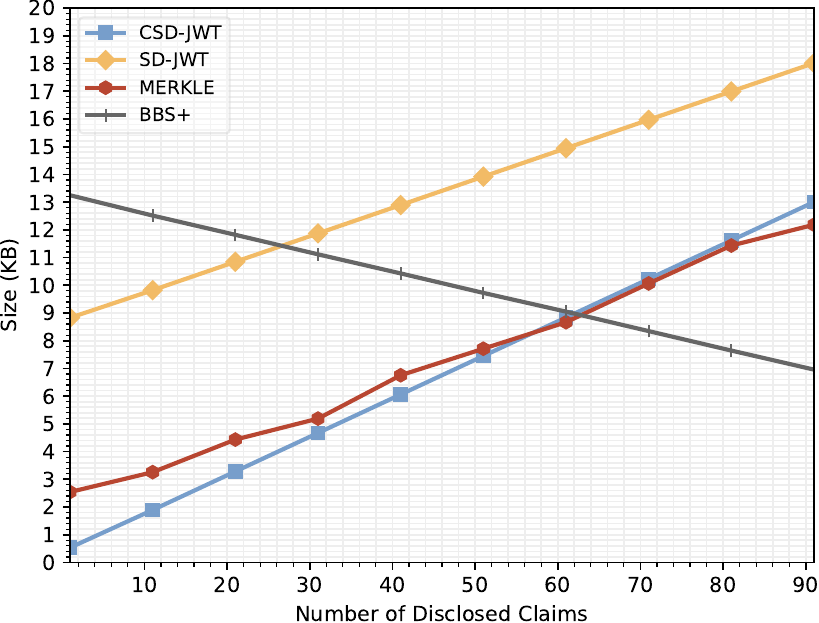}
    \caption{VP sizes (KB), based on a credential with $100$ claims.}
    \label{app:100_claims_vp_size}
\end{subfigure}
\caption{Comparison of \tool{}, SD-JWT, MTs, and BBS+ for VP size across different scenarios.}
\label{app:combined_vp_size}
\end{figure*}

\begin{figure*}[!t]
\centering
\begin{subfigure}{0.42\textwidth}
    \centering
    \includegraphics[width=\linewidth]{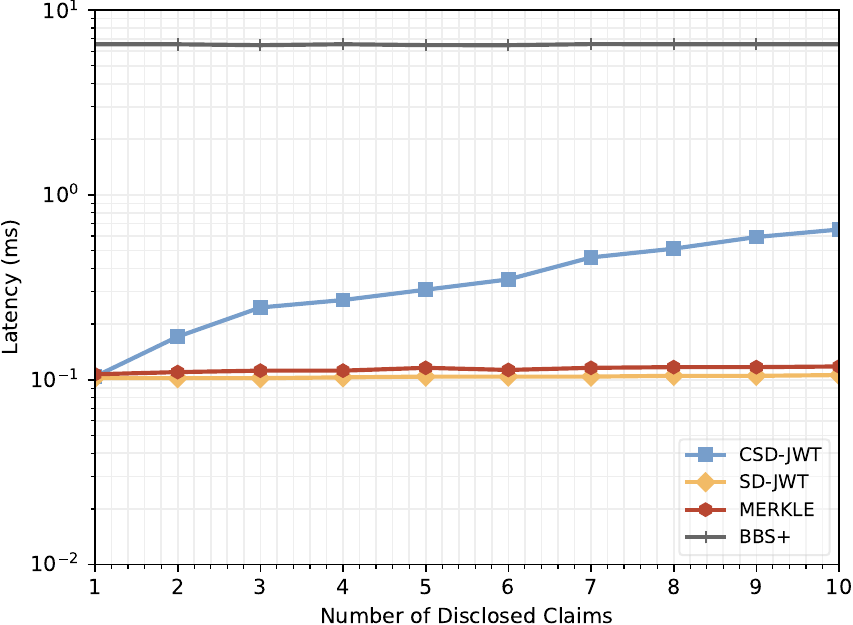}
    \caption{VP verification latency (ms) for credential with $10$ claims.}
    \label{app:10_claims_vp_verification_latency}
\end{subfigure}
\hfill
\begin{subfigure}{0.42\textwidth}
    \centering
    \includegraphics[width=\linewidth]{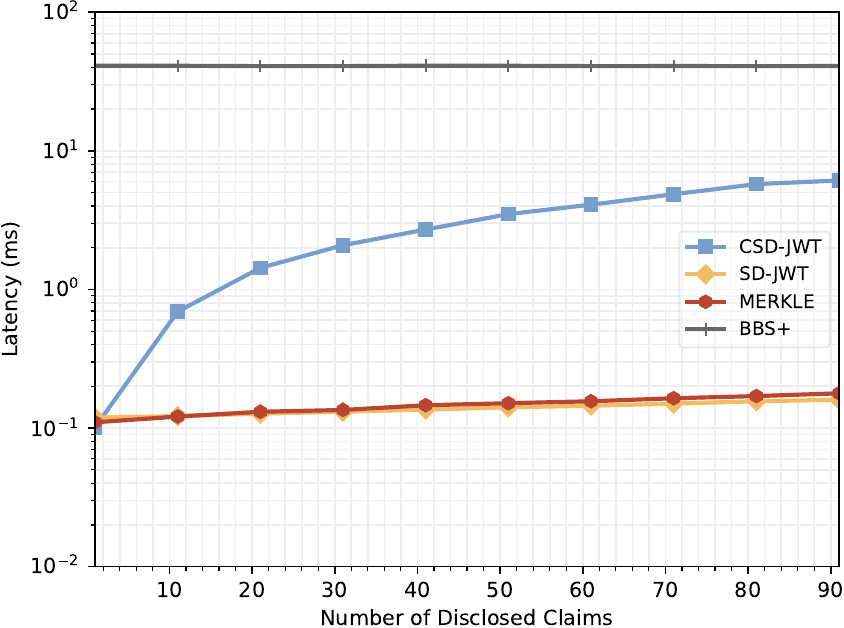}
    \caption{VP verification latency (ms) for credential with $100$ claims.}
    \label{app:100_claims_vp_verification_latency}
\end{subfigure}
\caption{Comparison of \tool{}, SD-JWT, MTs, and BBS+ for VP verification across different scenarios.}
\label{app:combined_vp_verification_latency}
\end{figure*}

\subsection{Generating and Verifying Verifiable Presentations}
In line with the main body of the paper, the following experiments evaluate the generation time and size of VPs derived from VCs with varying claim counts and disclosure levels. Specifically, we generated 10 VCs with claim counts ranging from 10 to 100 (in steps of 10), and for each, varied the percentage of disclosed claims from $0\%+1$ to $90\%+1$.

\smallskip
\noindent \textbf{Generation of Verifiable Presentations.}
We evaluated the latency of each selective disclosure mechanism for generating VPs. As shown in Figures~\ref{app:10_claims_vp_issuance_latency} and~\ref{app:100_claims_vp_issuance_latency}, both \tool{} and SD-JWT consistently deliver the fastest performance for credentials containing 10 and 100 claims. MT-based approaches follow closely, with slightly higher latency due to the overhead of tree construction. In contrast, BBS+ exhibits significantly higher computational cost, up to two orders of magnitude, primarily due to its reliance on complex cryptographic operations, which makes it unsuitable for constrained environments.

\smallskip
\noindent \textbf{Size of Verifiable Presentations.} 
Figure~\ref{app:combined_vp_size} presents the results of measuring the size of Verifiable Presentations VPs generated from Verifiable Credentials VCs containing 10 and 100 claims. As shown in Figure~\ref{app:10_claims_vp_size}, \tool{} consistently produces the most compact VPs across all disclosure levels for credentials with 10 claims.

However, when scaling to VCs with 100 claims, an interesting shift occurs. Figure~\ref{app:100_claims_vp_size} reveals an inflection point around 60 disclosed claims: below this threshold, \tool{} maintains its advantage in VP size, but beyond it, BBS+ begins to outperform. This behavior stems from the structural efficiency of BBS+ in handling large blocks of disclosed data. Unlike other mechanisms that encode each claim individually, BBS+ allows for compact proofs over multiple disclosed messages, which becomes increasingly beneficial as the number of disclosed claims grows.

\smallskip
\noindent \textbf{Verification of Verifiable Presentations.} Finally, we measured the time required for a verifier to validate the authenticity of VPs. As shown in Figures~\ref{app:10_claims_vp_verification_latency} and~\ref{app:100_claims_vp_verification_latency}, the results closely mirror those observed during VC issuance. SD-JWT and MT-based approaches remain the most efficient, with MTs incurring slightly higher overhead due to path verification. \tool{} consistently requires more computation during VP verification, while BBS+ performs the worst. This is primarily due to its reliance on cryptographic primitives, which are significantly more resource-intensive.

\subsection{Discussion of Results}
The results observed align with the computational nature of each mechanism. \tool{} demonstrates strong performance during VP generation, outperforming both MTs and BBS+. This advantage stems from its efficient witness filtering process in the WVC, which is comparable to salt filtering in SVCs and significantly less demanding than computing Merkle paths at runtime.

An additional insight emerges from Figure~\ref{app:combined_vp_size}, which compares VP sizes. BBS+ exhibits a unique trend: as more claims are disclosed, fewer ZKPs are needed to conceal the remainder, resulting in smaller network overhead. Conversely, the fluctuating behavior observed with MTs can be attributed to the benchmark’s structure, as when adjacent leaves are disclosed, the library optimizes proof generation by reducing the number of required nodes, even if the total number of proofs increases.
While BBS+ offers compact proofs in high-disclosure scenarios, its reliance on computationally intensive cryptographic operations for VP generation makes BBS+ impractical for constrained environments. Moreover, the VP size in many cases is still larger than \tool{} and MTs.

Among the evaluated methods, \tool{} offers a compelling trade-off between privacy and performance. While it may not be the optimal choice in every scenario, its efficient and lightweight operations make it especially suitable for resource-constrained environments, which is the central focus of this work.

}

\end{document}